\documentclass[12pt]{article}

\usepackage[totalwidth=467truept,totalheight=612truept]{geometry}
\usepackage{amsfonts,latexsym,amssymb,amsmath,graphicx,accents,float,slashed,subfigure}
\usepackage{dsfont}
\usepackage[T1]{fontenc}
\usepackage[hidelinks]{hyperref}

\global\arraycolsep=1truept

\numberwithin{equation}{section}

\begin{document}

\bigskip \phantom{C}

\vskip2truecm

\begin{center}
{\huge \textbf{Fakeons And Lee-Wick Models}}



\vskip1.8truecm

\textsl{Damiano Anselmi}

\vskip .2truecm

\textit{Dipartimento di Fisica ``Enrico Fermi'', Universit\`{a} di Pisa, }

\textit{Largo B. Pontecorvo 3, 56127 Pisa, Italy}

\textit{and INFN, Sezione di Pisa,}

\textit{Largo B. Pontecorvo 3, 56127 Pisa, Italy}

damiano.anselmi@unipi.it

\vskip1.8truecm

\textbf{Abstract}
\end{center}

The \textquotedblleft fakeon\textquotedblright\ is a fake degree of freedom,
i.e. a degree of freedom that does not belong to the physical spectrum, but
propagates inside the Feynman diagrams. Fakeons can be used to make
higher-derivative theories unitary. Moreover, they help us clarify how the
Lee-Wick models work. In this paper we study the fakeon models, that is to
say the theories that contain fake and physical degrees of freedom. We
formulate them by (nonanalytically) Wick rotating their Euclidean versions.
We investigate the properties of arbitrary Feynman diagrams and, among other
things, prove that the fakeon models are perturbatively unitary to all
orders. If standard power counting constraints are fulfilled, the models are
also renormalizable. The $S$ matrix is regionwise analytic. The amplitudes
can be continued from the Euclidean region to the other regions by means of
an unambiguous, but nonanalytic, operation, called average continuation. We
compute the average continuation of typical amplitudes in four, three and
two dimensions and show that its predictions agree with those of the
nonanalytic Wick rotation. By reconciling renormalizability and unitarity in
higher-derivative theories, the fakeon models are good candidates to explain
quantum gravity.

\vfill\eject

\tableofcontents

\section{Introduction}

\setcounter{equation}{0}\label{s0}

The Lee-Wick (LW) models are a subclass of higher-derivative theories, where
the free propagators contain complex conjugate pairs of extra poles, besides
the poles corresponding to the physical degrees of freedom and the degrees
of freedom due to the gauge fixing. The LW models are claimed to lead to a
perturbatively unitarity $S$ matrix \cite{leewick,LWqed,CLOP} due to a
certain compensation mechanism.

Various issues concerning the formulation of the LW theories remained open
for a long time. For example, if they are defined as initially suggested by
Lee \cite{lee}, the models violate Lorentz invariance \cite{nakanishi}. This
problem is due to the incompleteness of the initial Lee-Wick prescription.
Lee and Wick specified how to integrate on the loop energies, but did not
provide a compatible prescription for the integrals on the loop space
momenta.

To overcome these difficulties, further prescriptions were supplemented
later. For example, in ref. \cite{CLOP} a procedure of limit, which is known
as \textit{CLOP prescription}, was proposed to treat the critical situations
where the LW\ poles pinch the integration paths on the complex energy
planes. Lorentz invariance is recovered \cite{grinstein}, but in some
one-loop diagrams the CLOP prescription is ambiguous \cite{LWformulation}
and other ambiguities are expected at higher orders \cite{CLOP}. Moreover,
it is unclear how to incorporate the CLOP prescription at the Lagrangian
level or in the Feynman rules.

The problems were recently solved by reformulating the LW models by
(nonanalytically) Wick rotating their Euclidean versions \cite{LWformulation}%
. This procedure not only provides the correct prescription to integrate on
the loop energies, which agrees with the Lee-Wick one, but also provides the
natural companion prescription to integrate on the loop space momenta.

Briefly, the Lee-Wick integral on the loop energies includes complex values,
so an integral on real values of the loop space momenta is not compatible
with Lorentz invariance. However, if the integration domain on the loop
space momenta is deformed in a suitable way to include complex values,
Lorentz invariance is recovered.

It turns out that the Wick rotation is analytic only in a region of the
space $\mathcal{P}$ of the (complexified) external momenta, the region that
contains the purely imaginary energies. We call it \textit{main region} and
denote it by $\mathcal{A}_{0}$. The Wick rotation is nonanalytic elsewhere,
due to the LW pinching \cite{LWformulation}. In the end, the space $\mathcal{%
P}$ is divided into disjoint regions $\mathcal{A}_{i}$ of analyticity. A
loop integral gives an analytic function in each $\mathcal{A}_{i}$. The
relations among the functions associated with different regions are
unambiguous, but not analytic.

The domain deformation mentioned above is simple to formulate, but hard to
implement practically. Fortunately, there exists a shortcut to get directly
to the final result, which is simple and powerful. As said, the Wick
rotation is analytic in $\mathcal{A}_{0}$. The obstacles that prevent the
analytic continuation beyond $\mathcal{A}_{0}$ are the LW thresholds,
associated with LW poles that pinch the integration paths on the energies. The thresholds
have the form $p^{2}=\tilde{M}^{2}$, where $p$ is a linear combination of
incoming momenta and $\tilde{M}$ is a linear combination of (possibly
complex) masses. A LW threshold can be analytically overcome in two
independent ways. Neither of the two is separately compatible with unitarity
and there is no way to choose between them. We show that the nonanalytic
Wick rotation picks the arithmetic average of the two continuations, which
we call \textit{average continuation}. The final amplitudes are unitary,
Lorentz invariant and analytic in every $\mathcal{A}_{i}$, $i\neq 0$,
although not analytically related to the amplitudes evaluated in $\mathcal{A}%
_{0}$.

In this paper we study these issues in detail in arbitrary diagrams and show
that the formulation of the LW models is consistent to all orders. We
compute the average continuation of typical physical amplitudes in four,
three and two spacetime dimensions and provide numerical checks that the
average continuation and the nonanalytic Wick rotation give the same
results. Moreover, we prove that the LW models are perturbatively unitary to
all orders and show that their renormalization coincides with the one of
their Euclidean versions. This property ensures that the locality of
counterterms and the usual rules of power counting hold in every region $%
\mathcal{A}_{i}$.

The average continuation is an extremely powerful tool. It simplifies the
computation of the amplitudes in the regions $\mathcal{A}_{i}$, $i\neq 0$.
It eliminates the need of starting from the Euclidean space and performing the
Wick rotation. It allows us to prove the perturbative unitarity to all
orders in a relatively straightforward way. It gives an efficacious control
on the renormalization.

In ref. \cite{LWunitarity} the perturbative unitarity of the LW models was
proved at one loop. The generalization of the proof to all orders can be
worked out by first deriving the so-called cutting equations \cite%
{unitarity,unitaritymio,ACE} (which imply the unitarity equation $%
SS^{\dagger }=1$), in the main region $\mathcal{A}_{0}$ and then proving
that they can be average-continued to the other regions $\mathcal{A}_{i}$.
The final cutting equations have the expected, unitary form and propagate
only the physical degrees of freedom. We actually need to work with
generalized versions of the equations, which are proved starting from the 
\textit{algebraic cutting equations} (ACE) of ref. \cite{ACE}, a set of
polynomial identities associated with Feynman diagrams which are
particularly fit to perform the average continuation from $\mathcal{A}_{0}$
to $\mathcal{A}_{i}$.

We recall that the cutting equations imply $SS^{\dagger }=1$
straightforwardly in the models involving just scalar fields and fermions.
In gauge theories \cite{thooft} and gravity \cite{unitaritymio}, they imply
a pseudounitarity equation, which turns into the unitarity equation after
proving that the temporal and longitudinal components of the gauge fields
are compensated by the Faddeev-Popov ghosts.

It is important to stress that not all the higher-derivative theories fall
in the Lee-Wick class. For example, the Lee-Wick models of quantum gravity
are typically superrenormalizable. The reason is that the LW poles must come
in complex conjugate pairs, which requires many higher derivatives. With
fewer higher derivatives we may build a strictly renormalizable theory \cite%
{stelle}, but then the free propagators have ghost poles with real squared
masses. In ref. \cite{LWgravmio} it was shown that it is possible to double
such poles by means of a new quantization prescription and treat them as LW
poles associated with a fictitious LW scale $\mathcal{E}$ that is sent to
zero at the very end. This leads to the introduction of the notion of fake
degree of freedom, or \textquotedblleft fakeon\textquotedblright . Once a
pole is doubled according to this prescription, it can be consistently
dropped from the physical spectrum. Turning ghosts into fakeons allows us to
make the higher-derivative theories unitary.

The notion of fakeon generalizes the ideas of Lee and Wick and actually
clarifies their crucial properties. For example, the nonanalyticity of the $%
S $ matrix due to the LW pinching can be seen as associated with a fakeon of
a finite LW scale $\mathcal{E}=M$. For this reason, the LW models are
particular \textquotedblleft fakeon models\textquotedblright , by which we
mean models with physical degrees of freedom and fakeons. The results of
this paper, such as the proof of perturbative unitarity to all orders, hold
in all the fakeon models.

We recall that the LW\ models are also investigated for their possible
phenomenological implications, for example in QED \cite{LWqed}, the standard
model \cite{LWstandardM} and grand unified theories \cite{LWunification},
besides quantum gravity \cite{LWgrav,LWgrav2,LWgravmio}. The results of this
paper and refs. \cite{LWformulation,LWunitarity,LWgravmio} raise the fakeon
models to the status of consistent fundamental theories, since the
theoretical problems that could justify a certain skepticism around them are
now overcome. In particular, we have viable candidates to explain quantum
gravity within quantum field theory. Among the various possibilities, a
unique one is strictly renormalizable \cite{LWgravmio}.

The paper is organized as follows. In sections \ref{lws} and \ref{thedomdef}
we recall the formulation of the Lee-Wick models as nonanalytically Wick
rotated Euclidean theories and investigate their main properties in
arbitrary Feynman diagrams. In particular, in section \ref{lws} we study the
LW\ pinching, while in section \ref{thedomdef} we study the domain
deformation. In section \ref{avediffe} we define the average continuation of
an analytic function and analyse its properties. We also define the
difference continuation, which is useful for the cutting equations. In
section \ref{avedime} we study the average continuation of typical
amplitudes in various dimensions and numerically compare the results with
those of the nonanalytic Wick rotation. In section \ref{fakeons} we recall
the definition of fakeon and it main properties. In section \ref{uni} we
prove the perturbative unitarity of the fakeon models to all orders. In
section \ref{renormalization} we show that the counterterms of the fakeon
models are the same as those of their Euclidean versions. Section \ref%
{conclusions} contains the conclusions.

\section{Lee-Wick models}

\setcounter{equation}{0}\label{lws}

In this section we study the Lee-Wick models by nonanalytically Wick
rotating their Euclidean versions. The arguments hold to all orders in
spacetime dimensions $D$ greater than or equal to two, in local quantum
field theories whose free propagators have poles that are located
symmetrically with respect to the real axis of the complex energy plane,
with squared masses that have nonnegative real parts. The poles located on
the real axis are called \textit{standard poles} and the other ones are
called \textit{LW poles}. The standard poles are \textit{physical} if they
have positive residues.

\begin{figure}[t]
\begin{center}
\includegraphics[width=8truecm]{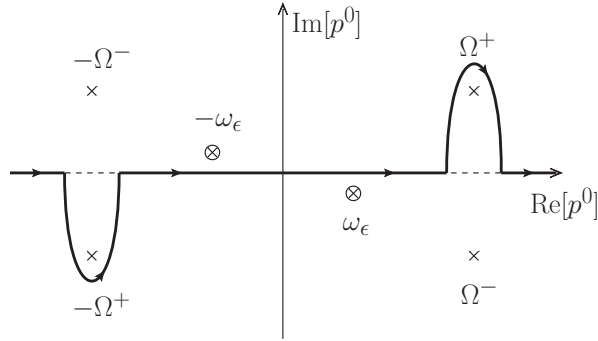}
\end{center}
\caption{Lee-Wick integration path}
\label{prop}
\end{figure}

Observe that derivative vertices and propagators with nontrivial numerators do not change
the analysis that follows. What matters in a loop integral are the
singularities of its integrand, i.e. the denominators of the propagators.

Before plunging into the nonanalytic Wick rotation, let
us stress why alternative approaches to the formulation of higher-derivative
theories are not viable. Letting aside \textit{ad hoc} prescriptions such as
the CLOP one, which cannot be incorporated at the level of the Feynman rules
and lead to ambiguous results, a natural formulation that may come to mind
is the Minkowski one, where the loop energies are integrated on their
natural, real values. Recently, it has been shown that the Minkowski
formulation generates nonlocal, non-Hermitian divergences that cannot be
removed by any standard procedures \cite{ugo}. In the few cases where the
locality of counterterms is not violated, the amplitudes are not consistent
with perturbative unitarity \cite{LWunitarity}. These observations lead us
to conclude that the Minkowski formulation is not the right one. The only
chance to define the higher-derivative models consistently is the Wick
rotation of their Euclidean versions.

The simplest example of LW propagator is%
\begin{equation}
S(p,m)=\frac{1}{p^{2}-m^{2}+i\epsilon }\frac{M^{4}}{(p^{2}-\mu
^{2})^{2}+M^{4}},  \label{propa}
\end{equation}%
where $M$ and $\mu $ are real mass scales. The poles of this propagator are shown in fig. \ref{prop}%
. The standard poles are encircled and read $p^{0}=\pm \omega _{\epsilon }(%
\mathbf{p})$, where $\omega _{\epsilon }(\mathbf{p})=\sqrt{\mathbf{p}%
^{2}+m^{2}-i\epsilon }$ and $p=(p^{0},\mathbf{p})$. The LW poles are not
encircled and read $p^{0}=\pm \Omega ^{+}(\mathbf{p})$ and $p^{0}=\pm \Omega
^{-}(\mathbf{p})$, where $\Omega ^{\pm }(\mathbf{p})=\sqrt{\mathbf{p}%
^{2}+M_{\pm }^{2}}$ and $M_{\pm }=\sqrt{\mu ^{2}\pm iM^{2}}$. We call the
pairs of poles $\Omega ^{\pm }$ and $-\Omega ^{\pm }$ \textit{Lee-Wick pairs}%
. Note that the Minkowski and Euclidean versions of the theories are not
equivalent, since the free propagators have poles in the first and third
quadrants of the complex plane.

Following ref. \cite{LWformulation}, the loop integrals are defined starting
from the Euclidean version of the theory. In the case of the tadpole
diagram, the Wick rotation leads to the integration path shown in fig. \ref%
{prop}. We see that the poles that are located to the right (resp. left) of
the imaginary axis are below (above) the integration path.

\begin{figure}[t]
\begin{center}
\includegraphics[width=14truecm]{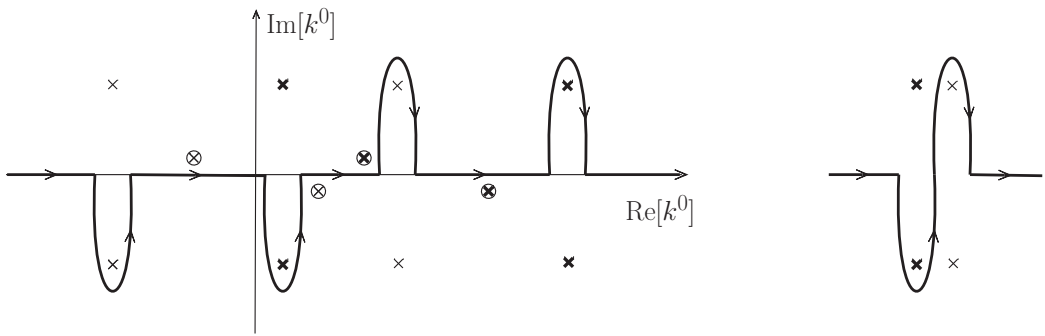}
\end{center}
\caption{Lee-Wick integration path for the bubble diagram (left) and LW
pinching (right)}
\label{WickBub2}
\end{figure}

The bubble diagram%
\begin{equation}
\mathcal{B}(p)=\int \frac{\mathrm{d}^{D}k}{(2\pi )^{D}}%
S(k,m_{1})S(k-p,m_{2}),  \label{bub}
\end{equation}%
which involves the product of two propagators, better illustrates the
general case. There, the Wick rotation leads to integration paths of the
form shown in the left picture of fig. \ref{WickBub2}. The thick crosses
denote the poles of the propagator $S(k-p,m_{2})$, which depend on $p$. The
other crosses denote the poles of $S(k,m_{1})$, which are fixed.

The general rule, which holds for arbitrary diagrams, is that the right
(resp. left) poles of a propagator -- i.e. those whose energies have
positive (negative) real parts at zero external momenta -- are located below
(above) the integration path.

When we wary $p$, a LW pole of $S(k-p,m_{2})$ can approach a LW pole of $%
S(k,m_{1})$ from the opposite side of the integration path. When the two come to coincide, we have a 
\textit{Lee-Wick\ pinching}. The standard poles can give the usual pinching,
which we call \textit{standard pinching}. Similarly, a \textit{mixed} LW pinching involves a LW pole and a standard pole.

\begin{figure}[t]
\begin{center}
\includegraphics[width=8truecm]{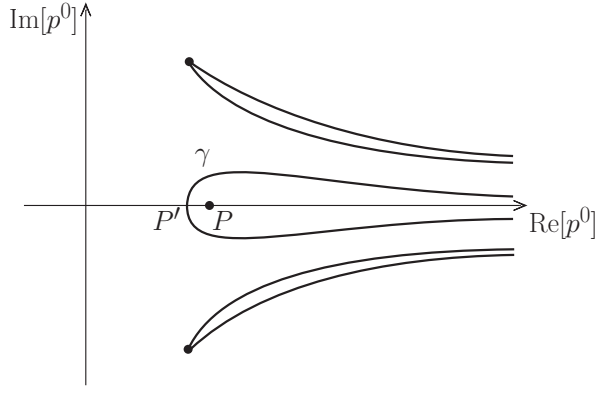}
\end{center}
\caption{Surfaces of LW\ pinching at $\mathbf{p}\neq 0$}
\label{completo}
\end{figure}

The condition for having a LW\ pinching is a system of two pole conditions.
For example, the right picture of fig. \ref{WickBub2} describes the
simultaneous pinching of the poles of two LW pairs. The conditions for the
top pinching are 
\begin{equation}
k^{0}=\Omega ^{+}(\mathbf{k}),\qquad k^{0}-p^{0}=-\Omega ^{-}(\mathbf{k-p}),
\label{toppi}
\end{equation}%
while the conditions for the bottom pinching are their complex conjugates
(with the understanding that the conjugation does not act on the momenta).
Solving (\ref{toppi}) for $k^{0}$, we obtain 
\begin{equation}
p^{0}=\Omega ^{+}(\mathbf{k})+\Omega ^{-}(\mathbf{k-p}).  \label{pinchcond}
\end{equation}%
Varying $\mathbf{k}$ in $\mathbb{R}^{3}$ with $\mathbf{p}$ real and fixed,
the solutions of this equation fill the region enclosed inside the curve $%
\gamma $ of fig. \ref{completo}.

Other LW pinchings occur for%
\begin{equation}
p^{0}=\Omega ^{+}(\mathbf{k})+\Omega ^{+}(\mathbf{k-p}),\qquad p^{0}=\Omega
^{-}(\mathbf{k})+\Omega ^{-}(\mathbf{k-p}),  \label{pc2}
\end{equation}%
and fill the regions enclosed inside the other two curves of fig. \ref%
{completo}. Finally, we have the regions obtained by reflecting (\ref%
{pinchcond}) and (\ref{pc2}) with respect to the imaginary axis.

Summarizing, the complex plane is divided into certain regions, which we
denote by $\mathcal{\tilde{A}}_{i}$. The curve $\gamma $ is the boundary of
the region $\mathcal{\tilde{A}}_{P}$ that intersects the\ positive real
axis. The region that contains the imaginary axis is denoted by $\mathcal{%
\tilde{A}}_{0}$.

The regions $\mathcal{\tilde{A}}_{i}$ are not Lorentz invariant, which is
the reason why they are not the final analytic regions $\mathcal{A}_{i}$.
For example, the threshold of the LW pinching given by eq. (\ref{pinchcond}) is the point $P$
with 
\begin{equation*}
p^{2}=2\mu ^{2}+2\sqrt{\mu ^{4}+M^{4}}\equiv 2M_{\text{LW}}^{2},
\end{equation*}%
as we prove below. However, the intersection between the curve $\gamma $ and the real axis is
not $P$, but a different point $P^{\prime }$. It is useful to introduce two
functions%
\begin{equation}
\eta _{\pm }(x)\equiv \frac{1}{\sqrt{2}}\sqrt{\sqrt{x^{2}+M^{4}}\pm x},
\label{etapm}
\end{equation}%
so that, for $x\geqslant 0$,%
\begin{equation*}
\sqrt{x\pm iM^{2}}=\eta _{+}(x)\pm i\eta _{-}(x).
\end{equation*}%
Then the point $P^{\prime }$ has energy 
\begin{equation}
p^{0}=2\eta _{+}(\mathbf{p}^{2}/4+\mu ^{2}).  \label{chip}
\end{equation}%
This relation cannot be expressed as a Lorentz invariant threshold condition
of the form $p^{2}=\tilde{M}^{2}$ for $\tilde{M}=2\eta _{+}(\mu ^{2})$.

For a while, we focus on real external momenta $p$, which are the ones of
physical interest. Note that (\ref{chip}) satisfies $4\mu ^{2}\leqslant
p^{2}\leqslant 2M_{\text{LW}}^{2}$, where the equalities holds for $\mathbf{p%
}^{2}=\infty $ and $\mathbf{p}=0$, respectively.

We define the Euclidean region as the strip $|\mathrm{Re}[p^{0}]|<|\mathbf{p}%
|$, which contains the imaginary axis. It is easy to check that the LW\
pinching conditions do not admit solutions there. Indeed, formulas (\ref%
{pinchcond}) and (\ref{pc2}) show that when a LW pinching occurs, the
minimum of $|\mathrm{Re}[p^{0}]|$ is the right-hand side of (\ref{chip}),
which is greater than or equal to $\sqrt{\mathbf{p}^{2}+4\mu ^{2}}$. In
particular, the Euclidean region is a subregion of $\mathcal{\tilde{A}}_{0}$.

We define the loop integral $\mathcal{B}(p)$ as follows. First, we integrate
on the loop energy $k^{0}$ by means of the residue theorem. Then, we
concentrate on the Euclidean region and integrate the loop space momentum $%
\mathbf{k}$ on its natural domain $\mathbb{R}^{3}$. Since no LW\ pinching
occurs, the result is analytic (and Lorentz invariant) but for the branch
cuts associated with the standard pinching.

Next, we ask ourselves if we can analytically extend the result away from
the Euclidean region. Focusing on the real axis, we find no obstacle for $%
p^{2}<4\mu ^{2}$, because all such points are below $P^{\prime }$. We can
also reach values $p^{2}\geqslant 4\mu ^{2}$, as long as we restrict the
Lorentz frame to the subset where the LW pinching does not occur for any $%
\mathbf{k}\in \mathbb{R}^{3}$. The good frames are those that have energies $%
p^{0}$ smaller than the energy of $P^{\prime }$. By formula (\ref{chip}),
this condition can be written as 
\begin{equation}
\qquad \mathbf{p}^{2}<\frac{4M^{4}}{p^{2}-4\mu ^{2}}-p^{2},  \label{pp}
\end{equation}%
which admits solutions if and only if $p^{2}<2M_{\text{LW}}^{2}$ (with $%
p^{2}\geqslant 4\mu ^{2}$).

In the end, for $p^{2}<2M_{\text{LW}}^{2}$, there is always an open subset $%
\mathcal{L}$ of Lorentz frames where no LW pinching occurs and we can
evaluate the loop integral by integrating $\mathbf{k}$ on $\mathbb{R}^{3}$.
The result is the analytic continuation of the function obtained in the
Euclidean region. Since it does not depend on the Lorentz frame, it can be
straightforwardly extended from $\mathcal{L}$ to the whole space of Lorentz
frames.

We have thus proved that the true LW threshold is the point $P$ of fig. \ref%
{completo}, beyond which the LW pinching is inevitable and the region $%
\mathcal{\tilde{A}}_{0}$ cannot be extended further. The region $\mathcal{A}%
_{0}$, which is the maximal extension of $\mathcal{\tilde{A}}_{0}$, stops at 
$P$.

The true challenge of the Lee-Wick models is to overcome the LW threshold $P$%
. To make a step forward towards the solution of this problem, we generalize
the calculation just described as follows. So far, we have calculated the
loop integral in a specific subset $\mathcal{L}$ of Lorentz frames, for $%
4\mu ^{2}<p^{2}<2M_{\text{LW}}^{2}$, because we wanted to be able to
integrate $\mathbf{k}$ on $\mathbb{R}^{3}$. Then, we extended the result to
all the Lorentz frames by Lorentz invariance. If we want to make the
calculation for $4\mu ^{2}<p^{2}<2M_{\text{LW}}^{2}$ directly in an
arbitrary Lorentz frame, we must deform the $\mathbf{k}$ integration domain $%
\mathcal{D}_{\mathbf{k}}$ to ensure that the LW pinching does not occur for
any $\mathbf{p}^{2}$. For example, if $\mathcal{O}_{P}$ denotes the portion
of the real axis with $p^{2}\geqslant 2M_{\text{LW}}^{2}$, $p^{0}>0$, we can
choose a deformation that squeezes the region $\mathcal{\tilde{A}}_{P}$ onto 
$\mathcal{O}_{P}$ (see the next section for details). Observe that $\mathcal{%
O}_{P}$ is Lorentz invariant.

The good news is that the domain deformation just mentioned allows us to
work out the loop integral even beyond the LW threshold $P$. In that case,
we have to proceed as follows. Let $\mathcal{\tilde{A}}_{P}^{\text{def}}$
denote the deformed region $\mathcal{\tilde{A}}_{P}$, before it is squeezed
onto $\mathcal{O}_{P}$. Let $\mathcal{D}_{\mathbf{k}}^{\text{def}}$ denote
the $\mathbf{k}$ integration domain associated with $\mathcal{\tilde{A}}%
_{P}^{\text{def}}$. We go inside $\mathcal{\tilde{A}}_{P}^{\text{def}}$ and
evaluate the loop integral $\mathcal{B}(p)$ there. Since the condition (\ref%
{pinchcond}) is complex, it can be split into two real conditions $x=y=0$
for suitable functions $x$ and $y$ of $\mathbf{k}$. Changing variables, in $%
D\geqslant 3$ the singularity has the form 
\begin{equation}
\frac{\mathrm{d}x\mathrm{d}y}{x+iy},  \label{integ}
\end{equation}%
which is integrable. In $D=2$ there is no singularity, because the pinching
just occurs at the boundaries $\gamma $, $\gamma ^{\text{def}}$ of the
regions $\mathcal{\tilde{A}}_{P}$, $\mathcal{\tilde{A}}_{P}^{\text{def}}$.
We view the result of the calculation in $\mathcal{\tilde{A}}_{P}^{\text{def}%
}$ as a function of the $\mathbf{k}$ integration domain $\mathcal{D}_{%
\mathbf{k}}^{\text{def}}$. When we finalize the deformation that takes $%
\mathcal{\tilde{A}}_{P}^{\text{def}}$ to $\mathcal{O}_{P}$, we obtain the
value of the loop integral on the real axis above the LW threshold $P$.

At the end, we can take $\mathcal{A}_{P}=\mathcal{O}_{P}$. Alternatively, we
can analytically continue the result found in $\mathcal{O}_{P}$ to a
neighborhood of $\mathcal{O}_{P}$ and take that neighborhood as the final
region $\mathcal{A}_{P}$ (reducing $\mathcal{A}_{0}$ correspondingly).

Before the squeezing of $\mathcal{\tilde{A}}_{P}\ $to $\mathcal{O}_{P}$, the
result of the loop integral in $\mathcal{\tilde{A}}_{P}$ is neither analytic
nor Lorentz invariant, in dimensions greater than or equal to three. On the
other hand, two dimensions are exceptional, because in $D=2$ the LW\
pinching occurs only at the boundaries of the regions $\mathcal{\tilde{A}}%
_{i}$, $i\neq 0$, but not inside. Consequently, the loop integral is both
Lorentz invariant and analytic in (a neighborhood of) $\mathcal{O}_{P}$,
even before making the domain deformation. We check these properties
explicitly in the examples of section \ref{avedime}. In the next section we
explain in detail how the domain deformation works in arbitrary $D\geqslant
2 $.

So far, we have focused on the LW thresholds that are located on the real
axis. Similar arguments hold for the other LW pinchings (\ref{pc2}), whose
thresholds are the points of minimum $\mathrm{Re}[p^{0}]$ of the
corresponding regions $\mathcal{\tilde{A}}_{i}$. It is easy to check that
such points have $\mathrm{Re}[p^{0}]=2\eta _{+}(\mathbf{p}^{2}/4+\mu ^{2})$
and $\mathrm{Im}[p^{0}]=\pm 2\eta _{-}(\mathbf{p}^{2}/4+\mu ^{2})$, so the
thresholds are $p^{2}=4\mu ^{2}\pm 4iM^{2}$. When $\mathbf{p}\rightarrow 0$
the corresponding regions $\mathcal{\tilde{A}}_{i}$ squeeze onto curves with
endpoints at the thresholds. The calculations beyond such thresholds are
performed with a procedure analogous to the one described above: first, we
evaluate the loop integral inside a region $\mathcal{\tilde{A}}_{i}$; then,
we deform the $\mathbf{k}$ integration domain till $\mathcal{\tilde{A}}_{i}$
gets squeezed onto a curve; finally, we take such a curve
as the final region $\mathcal{A}_{i}$, or enlarge it to some neighborhood of
it by analytic continuing the result found in it.

\begin{figure}[t]
\begin{center}
\includegraphics[width=10truecm]{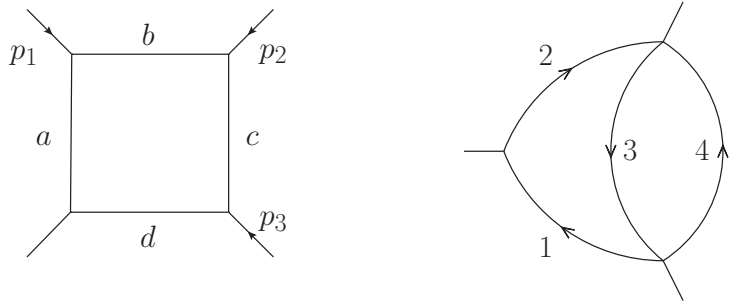}
\end{center}
\caption{Box and chestnut diagrams}
\label{chestnut}
\end{figure}

More complicated one-loop diagrams can be studied similarly. As an example,
consider the box diagram shown in the left picture of fig. \ref{chestnut}.
We assume that the propagators have the same masses $m$, $\mu $ and $M$, for
simplicity. The pinchings may occur when two, three or four propagators have
simultaneous pole singularities.

Decomposing into partial fractions, the integrand can be written as a sum of
terms 
\begin{equation}
\frac{1}{(z-\sigma _{1}a)(z-\sigma _{2}b)(z-\sigma _{3}c)(z-\sigma _{4}d)},
\label{abcd}
\end{equation}%
where $z$ denotes the loop energy $k^{0}$, $\sigma _{i}=\pm 1$ and each $%
a,b,c,d$ is a frequency $\omega _{\epsilon }$ or $\Omega ^{\pm }$ plus a
linear combination of incoming external energies. The poles with $\sigma
_{i}=1$ lie one side of the $z$ integration path, while the poles with $%
\sigma _{i}=-1$ lie on the other side. If all the $\sigma _{i}$ are equal,
the residue theorem gives zero. If $\sigma _{1}=1$, $\sigma _{2}=\sigma
_{3}=\sigma _{4}=-1$, the residue theorem give a result proportional to 
\begin{equation}
\frac{1}{(a+b)(a+c)(a+d)}.  \label{a1}
\end{equation}%
If $\sigma _{1}=\sigma _{3}=1$, $\sigma _{2}=\sigma _{4}=-1$, the residue
theorem gives%
\begin{equation}
\frac{1}{(a+d)(c+b)(c+d)}+\frac{1}{(a+b)(a+d)(c+b)}.  \label{a2}
\end{equation}%
Each singularity of (\ref{a1}) and (\ref{a2}) has the form (\ref{pinchcond})
or (\ref{pc2}). The other cases are permutations of the ones just
described. Note that the frequencies are always summed with positive
relative signs.

In the end, we only have situations that are analogous to those already met
in the case of the bubble diagram. The LW thresholds are%
\begin{equation*}
p_{i}^{2}=2M_{\text{LW}}^{2},\qquad (p_{1}+p_{2})^{2}=2M_{\text{LW}%
}^{2},\qquad (p_{2}+p_{3})^{2}=2M_{\text{LW}}^{2},\qquad
(p_{1}+p_{2}+p_{3})^{2}=2M_{\text{LW}}^{2},
\end{equation*}%
where $p_{i}$, $i=1,2,3$ denote the incoming momenta shown in the picture.

The evaluation of the loop integral proceeds as before. We first compute it
in the Euclidean region, where no LW pinching occurs, by integrating the
loop space momentum $\mathbf{k}$ on $\mathbb{R}^{3}$. Then we extend the
result by analytic continuation to $\mathcal{\tilde{A}}_{0}$. Third, we
maximize the region $\mathcal{\tilde{A}}_{0}$, again by analytic
continuation, which identifies the region $\mathcal{A}_{0}$. Beyond $%
\mathcal{A}_{0}$ we find obstacles, given by the LW thresholds. We overcome
those obstacles by going inside the regions $\mathcal{\tilde{A}}_{i}$, $%
i\neq 0$, and then deforming the $\mathbf{k}$ integration domain to squeeze
those regions into curves. At the end, we may define the
regions $\mathcal{A}_{i}$ as neighborhoods of those curves. We arrange each $\mathcal{A}_{i}$ so as to make it Lorentz invariant for real external momenta.

\subsection{LW pinching beyond one loop}

Before considering an arbitrary multiloop diagram, we begin with the
chestnut diagram shown in the right picture of fig. \ref{chestnut}. The
propagators 1 and 2 depend on one loop momentum, which we call $k$. The
integration path over $k^{0}$ gets pinched when two poles come to coincide
from opposite sides. This gives relations of the form 
\begin{equation*}
k^{0}=\tilde{\omega}_{1}(\mathbf{k}),\qquad k^{0}-p^{0}=-\tilde{\omega}_{2}(%
\mathbf{k-p}),
\end{equation*}%
where $\tilde{\omega}_{i}$ can stand for $\omega _{\epsilon }$ or $\Omega
^{\pm }$ and $p$ is an external momentum. Integrating over $k^{0}$ by means
of the residue theorem, we remain with a single pole, which occurs for 
\begin{equation}
p^{0}=\tilde{\omega}_{1}(\mathbf{k})+\tilde{\omega}_{2}(\mathbf{k-p}).
\label{c1}
\end{equation}%
This condition is analogous to (\ref{pinchcond}) and (\ref{pc2}).

Now, let us consider the propagators 1, 3 and 4. They depend on two loop
momenta, $k_{1}$ and $k_{2}$, which we assign to the legs 1 and 4. Their
simultaneous singularities give pinching conditions of the form%
\begin{equation*}
k_{1}^{0}=\tilde{\omega}_{1}(\mathbf{k}_{1}\mathbf{),\qquad }k_{2}^{0}=%
\tilde{\omega}_{2}(\mathbf{k}_{2}\mathbf{),}\qquad
k_{1}^{0}+k_{2}^{0}-p^{0}=-\tilde{\omega}_{3}(\mathbf{k}_{1}+\mathbf{\mathbf{%
k}}_{2}-\mathbf{p),}
\end{equation*}%
where $p$ is a sum of incoming external momenta. The signs in front of the
frequencies ensure that the first and third pole lie on opposite sides with
respect to the $k_{1}^{0}$ integration path, while the second and third pole
lie on opposite sides with respect to the $k_{2}^{0}$ integration path. The
integrals over $k_{1}^{0}$ and $k_{2}^{0}$ eliminate the first two
conditions and turn the third one into%
\begin{equation}
p^{0}=\tilde{\omega}_{1}(\mathbf{k}_{1})+\tilde{\omega}_{2}(\mathbf{k}_{2})+%
\tilde{\omega}_{3}(\mathbf{k}_{1}+\mathbf{\mathbf{k}}_{2}-\mathbf{p).}
\label{c2}
\end{equation}

Now, let us consider the contribution 
\begin{equation*}
\frac{1}{(z_{1}-a)(z_{1}+b)(z_{1}+z_{2}+c)(z_{2}-d)},
\end{equation*}%
where $z_{1}=k_{1}^{0}$, $z_{2}=k_{2}^{0}$ and $a$, $b$, $c$ and $d$ are
defined as before and associated with the legs 1, 2, 3, and 4, respectively.
A pinching can occur, since $a$ lies on one side of the $z_{1}$ integration
path, with $b$, $c$ on the other side of it, and at the same time $c$ and $d$
lie on opposite sides of the $z_{2}$ integration path. The residue theorem
gives a result proportional to%
\begin{equation*}
\frac{1}{(a+b)(a+c+d)}.
\end{equation*}%
The denominator vanishes in three situations, two \textit{minimal} and one
nonminimal. The minimal condition $a+b=0$ has the form (\ref{c1}). The
minimal condition $a+c+d=0$ has the form (\ref{c2}). The nonminimal
condition is the system made of the two.

The calculation can proceed as in the one loop case, the only difference
being that at some point we have to deform the integration domains of both
loop space momenta. The other contributions to the chestnut diagram can be
treated similarly.

The arguments just given can be generalized to diagrams with arbitrary
numbers of loops. The minimal configuration of pole singularities which may
give a pinching occurs when the number $n$ of propagators that have
simultaneous pole singularities is equal to the number of loop momenta they
depend on, plus one. If we parametrize the loop momenta in a convenient way,
the first $n-1$ conditions read $k_{i}^{0}=\tilde{\omega}_{i}(\mathbf{k}_{i}%
\mathbf{)}$, $i=1,\ldots n-1 $. After integrating on the loop energies $%
k_{i}^{0}$ by means of the residue theorem, the last condition becomes 
\begin{equation}
D_{\text{pinch}}=0,  \label{pinchgengen}
\end{equation}%
where%
\begin{equation}
D_{\text{pinch}}\equiv -p^{0}+\sum_{i=1}^{n-1}\tilde{\omega}_{i}(\mathbf{k}%
_{i}\mathbf{)}+\tilde{\omega}_{n}\left( \sum_{i=1}^{n}\mathbf{k}_{i}-\mathbf{%
p}\right)  \label{pinchgen}
\end{equation}%
and $p$ is again a sum of incoming external momenta. This is the minimal
pinching condition, with a convenient parametrization for the momenta. More
generally, the $k_{i}$ may be independent linear combinations of the loop
momenta (with coefficients $\pm 1$) plus linear combinations of the external
momenta.

The most general configuration of pole singularities arises as a
superposition of minimal configurations (plus configurations of
singularities that give no pinching, which we can ignore). Then, the most
general pinching condition is just a system made of minimal conditions. For
this reason, it is sufficient to study the minimal condition, in the
parametrization (\ref{pinchgen}).

We may have a pure LW pinching, where only LW poles are involved, a mixed LW
pinching, where both LW and standard poles are involved, and a standard
pinching, where only standard poles are involved.

An important fact is that the signs in front of the frequencies that appear
on the right-hand side of (\ref{pinchgen}) are always positive. The reason
is that the pinching just occurs between right and left poles of different
propagators, the right ones being placed below the integration path on the
loop energy and the left ones being placed above it. There is no pinching
between two right poles or two left poles (which would generate minus signs
in front of the frequencies), because they are located on the same side of
the integration path.

The threshold associated with the pinching condition (\ref{pinchgengen})-(%
\ref{pinchgen}) is%
\begin{equation}
p^{2}=\left[ \sum_{i=1}^{n}\tilde{\omega}_{i}(\mathbf{0)}\right] ^{2}.
\label{lwth}
\end{equation}%
This formula is a straightforward generalization of the one that holds in
the standard case, but must be proved anew, because the LW pinching involves
unusual features, such as the extended regions $\mathcal{\tilde{A}}_{i}$
that violate Lorentz invariance in some intermediate steps.

Specifically, the thresholds are found by means of a two-step procedure:
first we minimize $\mathrm{Re}[p^{0}]$ in $\mathbf{k}_{i}$ and then we
maximize $\mathrm{Re}[p^{2}]$ in $\mathbf{p}$. Referring to the analysis
made at one loop for thresholds on the real axis, the first step corresponds
to identifying the point $P^{\prime }$ of fig. \ref{completo} and the second
step corresponds to deforming $P^{\prime }$ into $P$. Now we prove that this
procedure does give formula (\ref{lwth}).

Let us first consider the case where only LW poles are involved, i.e. $n_{+}$
frequencies $\tilde{\omega}_{i}$ are equal to $\Omega ^{+}$ and $n_{-}$
frequencies $\tilde{\omega}_{i}$ are equal to $\Omega ^{-}$, with $%
n=n_{+}+n_{-}$. We have%
\begin{equation}
\mathrm{Re}[p^{0}]=\sum_{i=1}^{n-1}\eta _{+}(\mathbf{k}_{i}^{2}+\mu ^{2}%
\mathbf{)}+\eta _{+}\left( \mathbf{K}^{2}+\mu ^{2}\right) ,  \label{rep0}
\end{equation}%
where $\eta _{+}$ is defined in formula (\ref{etapm}) and%
\begin{equation*}
\mathbf{K}=\sum_{i=1}^{n-1}\mathbf{k}_{i}-\mathbf{p.}
\end{equation*}

Minimizing $\mathrm{Re}[p^{0}]$ in $\mathbf{k}_{i}$, we obtain $\mathbf{k}%
_{i}=\mathbf{p}/n$ for every $i$, which gives 
\begin{subequations}
\begin{eqnarray}
p^{0} &=&n\eta _{+}(\mathbf{p}^{2}/n^{2}+\mu ^{2}\mathbf{)}%
+i(n_{+}-n_{-})\eta _{-}(\mathbf{p}^{2}/n^{2}+\mu ^{2}\mathbf{),}
\label{pzero} \\
p^{2} &=&4n_{+}n_{-}\eta _{-}^{2}(\mathbf{p}^{2}/n^{2}+\mu ^{2}\mathbf{)}%
+n^{2}\mu ^{2}+i(n_{+}^{2}-n_{-}^{2})M^{2}.  \label{pzerob}
\end{eqnarray}%
The maximum of $\mathrm{Re}[p^{2}]$ in $\mathbf{p}$ is its value for $%
\mathbf{p}=0$, which gives the thresholds 
\end{subequations}
\begin{equation*}
p^{2}=(n_{+}^{2}+n_{-}^{2})\mu ^{2}+2n_{+}n_{-}\sqrt{\mu ^{2}+M^{4}}%
+i(n_{+}^{2}-n_{-}^{2})M^{2}.
\end{equation*}%
The result agrees with (\ref{lwth}), since $\Omega ^{\pm }(\mathbf{0})=\eta
_{+}(\mu ^{2})\pm i\eta _{-}(\mu ^{2})$. The thresholds on the real axis are
those with $n_{+}=n_{-}$.

Observe that\ no LW\ pinching occurs in the Euclidean region $|\mathrm{Re}%
[p^{0}]|<|\mathbf{p}|$. Indeed, using formula (\ref{pzero}) we find that
wherever a LW\ pinching occurs the inequalities 
\begin{equation*}
|\mathrm{Re}[p^{0}]|\geqslant |\mathrm{Re}[p^{0}]|_{\min }=n\eta _{+}(%
\mathbf{p}^{2}/n^{2}+\mu ^{2}\mathbf{)\geqslant }\sqrt{\mathbf{p}%
^{2}+n^{2}\mu ^{2}}
\end{equation*}%
hold.

Next, let us consider the mixed LW pinching, where both standard poles and
LW poles are present. We assume that $\mu $ and $M$ are the same everywhere,
but the standard masses are generic. We separate the last standard pole,
with mass $m$, from the other ones, with masses $m_{j}$ and loop space
momenta $\mathbf{q}_{j}$. Then, we get the condition (\ref{pinchgengen})
with 
\begin{equation}
D_{\text{pinch}}=-p^{0}+\sum_{i=1}^{n_{+}}\Omega ^{+}(\mathbf{k}_{i}\mathbf{%
)+}\sum_{i=n_{+}+1}^{n}\Omega ^{-}(\mathbf{k}_{i}\mathbf{)}%
+\sum_{j=1}^{r-1}\omega (\mathbf{q}_{j},m_{j})+\omega \left( \mathbf{Q,}%
m\right) .  \label{deno}
\end{equation}%
Here we have defined $\omega (\mathbf{p},m)=\sqrt{\mathbf{p}^{2}+m^{2}}$ and%
\begin{equation*}
\mathbf{Q}=\sum_{j=1}^{r-1}\mathbf{q}_{j}+\mathbf{K},\qquad \mathbf{K}%
=\sum_{i=1}^{n}\mathbf{k}_{i}-\mathbf{p}.
\end{equation*}%
First, we minimize $\mathrm{Re}[p^{0}]$ in $\mathbf{q}$, which is
straightforward. Indeed, translating $\mathrm{Re}[p^{0}]$ by a constant,
this operation just gives the threshold of the standard pinching. We thus
find 
\begin{equation}
\mathbf{q}_{j}=-\frac{m_{j}}{m_{\text{tot}}}\mathbf{K},\qquad m_{\text{tot}%
}=m+\sum_{j=1}^{r-1}m_{j},  \label{qj}
\end{equation}%
and%
\begin{equation*}
\mathrm{Re}[p^{0}]=\sum_{i=1}^{n}\eta _{+}(\mathbf{k}_{i}^{2}+\mu ^{2}%
\mathbf{)}+\omega (\mathbf{K},m_{\text{tot}}),
\end{equation*}

Now we minimize $\mathrm{Re}[p^{0}]$ in $\mathbf{k}_{i}$, which gives $%
\mathbf{k}_{i}=\mathbf{p}\alpha (\mathbf{p})\equiv \mathbf{s}$ for every $i$%
, for some function $\alpha $ of $\mathbf{p}$. It is convenient to express
everything in terms of $\mathbf{s}$ rather than $\mathbf{p}$. We find 
\begin{equation}
\mathbf{p}=n\mathbf{s}+\frac{2\mathbf{s}m_{\text{tot}}\eta _{+}^{\prime }}{%
\sqrt{1-4\mathbf{s}^{2}\eta _{+}^{\prime \hspace{0.01in}2}}},  \label{ppp}
\end{equation}%
where $\eta _{+}^{\prime }(x)=\mathrm{d}\eta _{+}^{\prime }(x)/\mathrm{d}x$.
Unless specified differently, here and below the arguments of $\eta _{+}$, $%
\eta _{-}$ and their derivatives are $\mathbf{s}^{2}+\mu ^{2}$. It is easy
to check that the argument of the square root in (\ref{ppp}) is always
positive.

Formula (\ref{qj}) gives $\mathbf{q}_{j}=m_{j}(\mathbf{p}-n\mathbf{s})/m_{%
\text{tot}}$. Using (\ref{ppp}) inside $D_{\text{pinch}}=0$, we get 
\begin{subequations}
\begin{eqnarray}
p^{0} &=&n\eta _{+}+\frac{m_{\text{tot}}}{\sqrt{1-4\mathbf{s}^{2}\eta
_{+}^{\prime \hspace{0.01in}2}}}+i(n_{+}-n_{-})\eta _{-},  \label{pizero} \\
\mathrm{Re}[p^{2}] &=&m_{\text{tot}}^{2}+n\mu ^{2}+4n_{+}n_{-}\eta
_{-}^{2}+2nm_{\text{tot}}\frac{\eta _{+}-2\mathbf{s}^{2}\eta _{+}^{\prime }}{%
\sqrt{1-4\mathbf{s}^{2}\eta _{+}^{\prime \hspace{0.01in}2}}}.  \label{rep2}
\end{eqnarray}%
At this point, we maximize $\mathrm{Re}[p^{2}]$ in $\mathbf{p}$. We can
actually maximize it in $\mathbf{s}$, since \textrm{d}$\mathbf{p}^{2}/%
\mathrm{d}\mathbf{s}^{2}$ is always positive. It is easy to show that the
right-hand side of (\ref{rep2}) is a monotonically decreasing function of $%
\mathbf{s}^{2}$, so the maximum of $\mathrm{Re}[p^{2}]$ coincides with its
value at $\mathbf{s}=0$, which gives the threshold

\end{subequations}
\begin{equation}
p^{2}=\left( n_{+}M_{+}+n_{-}M_{-}+m_{\text{tot}}\right) ^{2},  \label{pn}
\end{equation}%
in agreement with (\ref{lwth}).

Again, no LW\ pinching occurs in the Euclidean region $|\mathrm{Re}[p^{0}]|<|%
\mathbf{p}|$. Indeed, for arbitrary $\mathbf{k}_{i}$ and $\mathbf{q}_{j}$,
the LW\ pinching conditions $D_{\text{pinch}}=0$ imply 
\begin{equation*}
(\mathrm{Re}[p^{0}])^{2}-\mathbf{p}^{2}\geqslant (\mathrm{Re}[\tilde{p}%
^{0}])^{2}-\mathbf{p}^{2}\geqslant (\mathrm{Re}[\tilde{p}^{0}])^{2}-(\mathrm{%
\mathrm{Im}}[\tilde{p}^{0}])^{2}-\mathbf{p}^{2}=\mathrm{Re}[\tilde{p}%
^{2}]\geqslant 0,
\end{equation*}%
where $\tilde{p}=(\tilde{p}^{0},\mathbf{p})$ is the momentum $p$ that
minimizes $\mathrm{Re}[p^{0}]$ in $\mathbf{k}_{i}$ and $\mathbf{q}_{j}$,
encoded in formulas (\ref{pizero}) and (\ref{rep2}).

Consider a Feynman diagram $G$ with $n+1$ external legs. Let $p_{1},\cdots
,p_{n}$ denote the incoming momenta of $n$ external legs. The thresholds
read 
\begin{equation}
\left( \sum_{i\in I}p_{i}\right) ^{2}=\tilde{M}^{2},  \label{thresh}
\end{equation}%
where $I$ is a subset of indices of the incoming momenta and $\tilde{M}$ is
positive sum of ordinary masses $m$ and LW masses $M_{\pm }$. Note that the
incoming momentum of the $(n+1)$th external leg is%
\begin{equation*}
p_{n+1}=-\sum_{i=1}^{n}p_{i},
\end{equation*}%
so whenever a sum of incoming momenta includes $p_{n+1}$ it can be written
as minus a sum of $p_{i}$. Since the overall sign is immaterial for the
left-hand side of (\ref{thresh}), we can always write the thresholds as in
that formula.

\begin{figure}[t]
\begin{center}
\includegraphics[width=10cm]{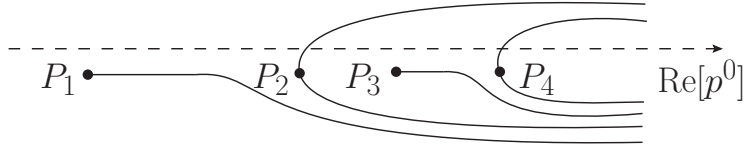}
\end{center}
\caption{Standard and LW thresholds close to the real axis}
\label{mixing}
\end{figure}

The number of thresholds (\ref{thresh}) and regions $\mathcal{\tilde{A}}_{i}$
of each loop integral is finite. If the masses $m_{j}$ are nonvanishing and
finitely many, the number of thresholds of an
amplitude is finite within any compact energy range, even after summing the
loop corrections to all orders. That number becomes infinite when some
masses $m_{j}$ vanish. This is the known problem of the infrared
divergences, which is dealt with by means of resummation techniques \cite%
{infred}.

Strictly speaking, the sum $m_{\text{tot}}$ of standard masses in formula (%
\ref{pn}) should be equipped with a small negative imaginary part, coming
from the width $\epsilon $ of the propagator (\ref{propa}). In several
calculations, as well as the proof of perturbative unitarity of section \ref%
{uni}, it is necessary to work at $\epsilon \neq 0$. Then the thresholds (%
\ref{pn}) with $n_{+}=n_{-}$ are not exactly on the real axis for $m_{\text{%
tot}}\neq 0$, but a bit displaced from it. As before, when LW poles are
involved, the conditions (\ref{pinchgengen}) identify extended regions $%
\mathcal{\tilde{A}}_{i}$, $i\neq 0$. Since $\epsilon $ is supposed to be
small, while $M$ is finite, the regions $\mathcal{\tilde{A}}_{i}$ always
intersect the real axis in a segment, when $n_{+}=n_{-}$. A typical
situation is shown in fig. \ref{mixing}, where $P_{1}$ and $P_{3}$ are
standard thresholds, while $P_{2}$ and $P_{4}$ are LW thresholds. For
convenience, we have drawn the branch cuts ending at the standard thresholds
so that they do not intersect the regions $\mathcal{\tilde{A}}_{i}$, $i\neq
0 $.

A loop integral $\mathfrak{I}$ is first evaluated in the Euclidean region,
by integrating on the natural real domain $\mathbb{R}^{3(n+r-1)}$ of the
loop space momenta $\mathbf{k}_{i}$ and $\mathbf{q}_{j}$. Then the result is
extended by analytic continuation to $\mathcal{\tilde{A}}_{0}$ and $\mathcal{%
A}_{0}$. Above the LW\ thresholds, the integration domain $\mathcal{D}_{%
\mathbf{k},\mathbf{q}}$ on $\mathbf{k}_{i}$ and $\mathbf{q}_{j}$ is deformed
from $\mathbb{R}^{3(n+r-1)}$ till the regions $\mathcal{\tilde{A}}_{i}$, $%
i\neq 0$, squeeze onto Lorentz invariant surfaces $\mathcal{L}_{i}$. The
calculation of $\mathfrak{I}$ is performed inside each deformed $\mathcal{%
\tilde{A}}_{i}$, $i\neq 0$, before finalizing the squeezing. Once the
squeezing is finalized, the results found in the surfaces $\mathcal{L}_{i}$
are extended to neighborhoods of them by analytic continuation. Those
neighborhoods can be taken as the the regions $\mathcal{A}_{i}$, $i\neq 0$.
For every threshold with $n_{+}=n_{-}$, the corresponding region $\mathcal{A}%
_{i}$ is enlarged enough till it intersects the real axis in a segment, as
in fig. \ref{mixing}. Note that the singularities $1/D_{\text{pinch}}$
associated with the LW pinchings have the form (\ref{integ}) and so are
integrable.

\section{The domain deformation}

\label{thedomdef}

\begin{figure}[t]
\begin{center}
\includegraphics[width=14truecm]{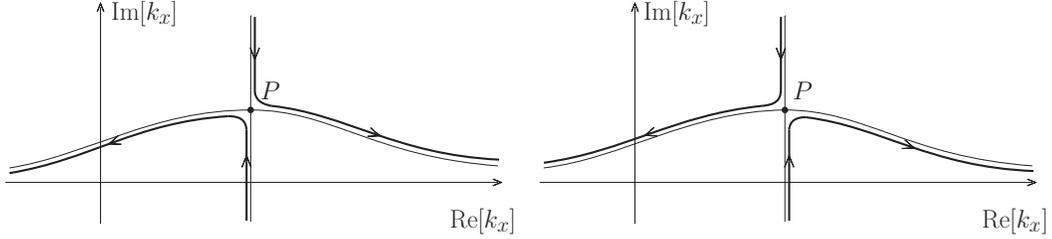}
\end{center}
\caption{Solutions (\protect\ref{kx}) of the pinching condition (\protect\ref%
{pinchcond}) as functions of $p^{0}$ in $D=2$, for Im$[p^{0}]\gtrsim 0$ and
Im$[p^{0}]\lesssim 0$, respectively. The vertical line is Re$[k_{x}]=p_{x}/2$%
}
\label{domdefR}
\end{figure}

In the most general case, the deformation of the integration domain on the
loop space momenta, required by the nonanalytic Wick rotation, is a rather
involved process. However, its main features are relatively simple. In this
section we illustrate them in detail, starting from the bubble diagram in $%
D=2$, then generalizing the arguments to arbitrary $D$ and arbitrary
diagrams.

\subsection{Domain deformation in the bubble diagram}

\begin{figure}[t]
\begin{center}
\includegraphics[width=10truecm]{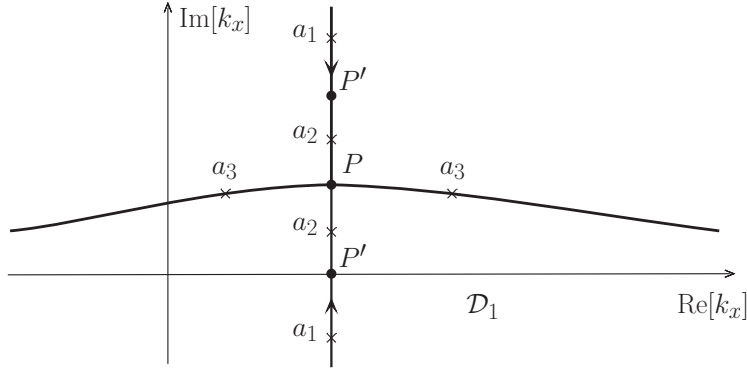}
\end{center}
\caption{Solutions (\protect\ref{kx}) for Im$[p^{0}]=0$}
\label{domdef}
\end{figure}

Consider the LW pinching condition (\ref{pinchcond}) in $D=2$, setting $%
p=(p^{0},p_{x})$, $k=(k^{0},k_{x})$. The solutions $k_{x}$ read

\begin{equation}
k_{x}^{\pm }(p)=\frac{p_{x}}{2}+i\frac{M^{2}}{p^{2}}\left( p_{x}\pm p^{0}%
\sqrt{1+\frac{\mu ^{2}}{M^{4}}p^{2}-\frac{(p^{2})^{2}}{4M^{4}}}\right) .
\label{kx}
\end{equation}%
Let us keep $p_{x}$ fixed (and real) and view $k_{x}^{\pm }$ as functions of 
$p^{0}$. If we move $p^{0}$ on fig. \ref{completo} along lines parallel to
the real axis, with Im$[p^{0}]\gtrsim 0$ and Im$[p^{0}]\lesssim 0$, we
obtain the pictures of fig. \ref{domdefR} (where $M=\mu =p_{x}=1$). In each
picture, the trajectories are the functions $k_{x}^{\pm }(p)$ and the arrows
point towards growing values of Re$[p^{0}]$. As long as Im$[p^{0}]\neq 0$,
the trajectories do not intersect each other. If we take the limit Im$%
[p^{0}]\rightarrow 0$, we obtain fig. \ref{domdef}, where the points $a_{i}$
with the same index $i$ correspond to solutions $k_{x}$ with the same value
of $p^{0}$. 

\begin{figure}[t]
\begin{center}
\includegraphics[width=16truecm]{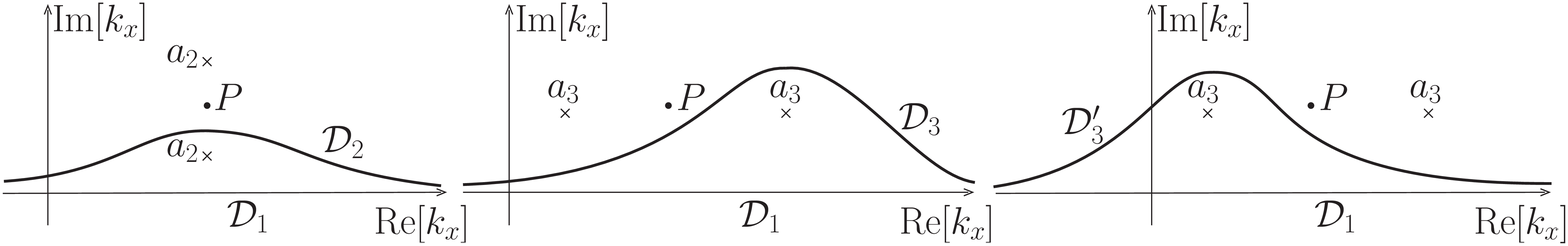}
\end{center}
\caption{Basic domain deformation}
\label{domdeffa}
\end{figure}

The natural $k_{x}$ integration domain is the $k_{x}$ real axis. In this
discussion we denote it by $\mathcal{D}_{1}$. Let us follow the solutions of
fig. \ref{domdef} and see how the integration domain must be deformed to
have analyticity. Referring to fig. \ref{completo}, we start from the
segment of the $p^{0}$ real axis that is located below $P^{\prime }$. A
typical point there is sent into the two points $a_{1}$ of fig. \ref{domdef}%
, which are located on opposite sides of the domain $\mathcal{D}_{1}$. 
When $p^{0}$ increases, one
trajectory $k_{x}^{\pm }$ intersects $\mathcal{D}_{1}$ (which happends when $%
p^{0}$ reaches the point $P^{\prime }$). The segment of the $p^{0}$ real
axis contained between $P^{\prime }$ and $P$ is represented by the points $%
a_{2}$ in the $k_{x}$ complex plane, which are located on the same side of $%
\mathcal{D}_{1}$. The loop integral, as a function of $p^{0}$, can be
analytically extended beyond $P^{\prime }$ by deforming $\mathcal{D}_{1}$
into some new domain $\mathcal{D}_{2}$ that looks like the one shown in the
first picture of fig. \ref{domdeffa}, so that the points $a_{2}$ are
left on opposite sides.

When $p^{0}$ continues to increase and reaches the point $P$, the two
trajectories hit each other. There, it is impossible to keep the solutions $%
k_{x}^{\pm }$ on opposite sides of the $k_{x}$ integration domain. This
means that the loop integral cannot be analytically extended beyond $P$ by
moving $p^{0}$ along the real axis. The point $P$ is the sole and true case
where the pinching cannot be avoided. It is obtained by setting the argument
of the square root of (\ref{kx}) to zero, which gives the LW threshold $%
p^{2}=2M_{\text{LW}}^{2}$, in agreement with the results of the previous
section.

Larger real values of $p^{0}$ take us into the portion $\mathcal{O}_{P}$ of
the real axis above $P$, which is represented by the points $a_{3}$ of fig. %
\ref{domdef}. There are two types $\mathcal{D}_{3}$ and $\mathcal{D}%
_{3}^{\prime }$ of deformed domains that leave those points on opposite
sides, as shown in the second and third pictures of fig. \ref{domdeffa}. The two
possibilities correspond to reaching $\mathcal{O}_{P}$ by giving $p^{0}$ a
small positive, or small negative, imaginary part. Indeed, we know from fig. %
\ref{domdefR} that the analytic continuation finds no obstacles in those
cases, because the $k_{x}$ trajectories never intersect each other.

In the end, we have two analytic continuations from $\mathcal{\tilde{A}}_{0}$
to $\mathcal{O}_{P}$, one obtained by circumventing $P$ from the half plane
Im$[p^{0}]>0$ and the other one obtained by circumventing $P$ from the half
plane Im$[p^{0}]<0$. We will see in section \ref{avediffe} that the result
of the loop integral above $P$ is the arithmetic average of the two (average
continuation).

\begin{figure}[t]
\begin{center}
\includegraphics[width=9truecm]{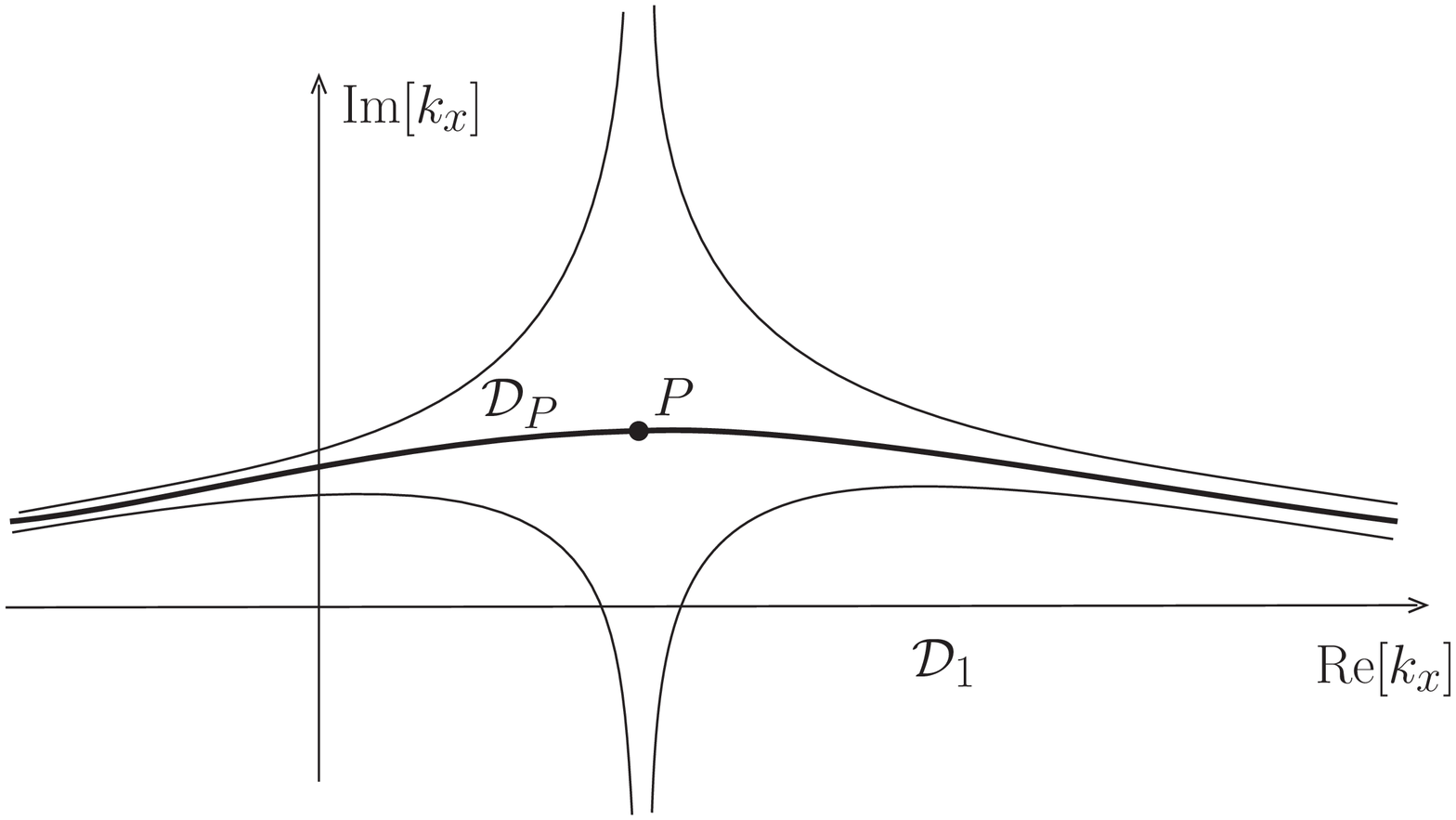}
\end{center}
\caption{Domain $\mathcal{D}_{P}$ that squeezes $\mathcal{\tilde{A}}_{P}$
onto $\mathcal{O}_{P}$}
\label{Dp}
\end{figure}

Finally, the region $\mathcal{\tilde{A}}_{P}$ can be completely squeezed
onto $\mathcal{O}_{P}$ by deforming $\mathcal{D}_{1}$ into the domain $%
\mathcal{D}_{P}$ made of the curve that crosses the points $a_{3}$ of fig. %
\ref{domdef}. Indeed, fig. \ref{Dp} shows that $\mathcal{D}_{P}$ always
leaves the solutions $k_{x}^{\pm }$ on the same side, no matter how small $|$%
Im$[p^{0}]|$ is taken.

The arguments can be easily extended to arbitrary dimensions $D$ greater
than two. Assume that the external space momentum $\mathbf{p}$ is directed
along the $x$ direction. Writing $p=(p^{0},p_{x},\mathbf{0})$ and $%
k=(k^{0},k_{x},\mathbf{k}_{\perp })$, it is easy to check that the
conditions (\ref{pinchcond}) and (\ref{pc2}) in $D>2$ are obtained from
those in $D=2$ by means of the replacement $\mu ^{2}\rightarrow \mathbf{k}%
_{\perp }^{2}+\mu ^{2}\equiv \tilde{\mu}^{2}$.\ Then it is apparent that to
squeeze the region $\mathcal{\tilde{A}}_{P}$ onto $\mathcal{O}_{P}$ we do
not need to deform $\mathbf{k}_{\perp }$ to complex values, since it is
enough to deform the $k_{x}$ integration domain as explained above, for
every $\tilde{\mu}^{2}$.

To summarize, the equations (\ref{pinchcond}) and (\ref{pc2}) tell us when
the integration path on the loop energy gets pinched. However, in most cases
the pinching is eventually avoided by deforming the integration domain on
the loop space momenta. The pinching is inevitable only at the LW
thresholds. Since the LW thresholds are Lorentz invariant, Lorentz
invariance is never truly violated. Moreover, the regions $\mathcal{\tilde{A}%
}_{i}$, $i\neq 0$, can be deformed and squeezed at will. The regions located
above the LW thresholds can be reached analytically from the regions located
below the LW thresholds in two independent ways.

It should also be noted that everything we have said so far equally applies
to the standard thresholds and actually offers a new approach to investigate
their properties. In the limit $M\rightarrow 0$ the solutions (\ref{kx})
become%
\begin{equation*}
k_{x}^{\pm }(p)=\frac{p_{x}}{2}\pm \frac{p^{0}}{2}\sqrt{1-\frac{4\mu ^{2}}{%
p^{2}}}.
\end{equation*}

In particular, we can appreciate why the thresholds are the only points of
true pinching, while the points lying on the branch cuts are not. Indeed,
the branch cuts can be displaced at will by deforming the integration
domains on the loop space momenta.

\subsection{Domain deformation in more complicated diagrams}

Now we study the domain deformation in the diagrams with more loops and/or
more independent external momenta.

If we have a single threshold, the analysis of the previous subsection can
be repeated with straightforward modifications. A unique combination $p$ of
external momenta is involved. If the pinching conditions involve a unique
loop momentum $k$, the analysis is exactly the same as before. If they
involve more than one loop momenta, we simply have more freedom to perform
the domain deformation.

\begin{figure}[t]
\begin{center}
\includegraphics[width=9cm]{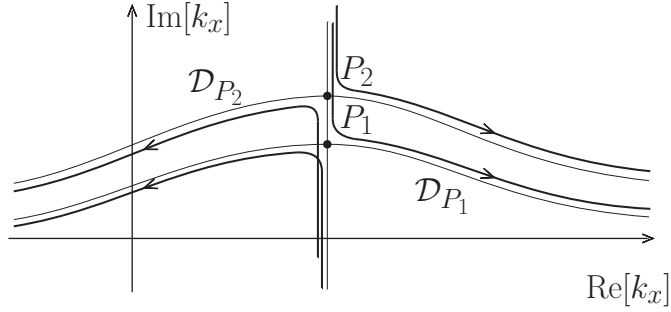}
\end{center}
\caption{Multiple thresholds with the same external momentum $p$}
\label{domdef2loops}
\end{figure}

Thus, we can concentrate on the case of multiple LW\ thresholds. We begin
from two LW thresholds involving the same combination $p$ of external
momenta. We denote them by $P_{1}$: $p^{2}=\tilde{M}_{1}^{2}$, $P_{2}$: $%
p^{2}=\tilde{M}_{2}^{2}$, etc. Let us assume the worst scenario (which
reduces the freedom to make the deformation to a minimum), where there is a
unique loop momentum $k$. As before, we choose $p=(p^{0},p_{x},\mathbf{0})$, 
$k=(k^{0},k_{x},\mathbf{k}_{\perp })$. Consider the condition (\ref%
{pinchcond}) with $\mu \rightarrow \mu _{1}$, $M\rightarrow M_{1}$, together
with the same condition (\ref{pinchcond}) with $\mu \rightarrow \mu _{2}$, $%
M\rightarrow M_{2}$. Formula (\ref{kx}), suitably adapted to the present
case, gives the solutions $k_{x}$. Observe that the vertical line Re$%
[k_{x}]=p_{x}/2$ of fig. \ref{domdef} does not depend on the masses and $\mathbf{%
k}_{\perp }$, so it is the same for every threshold. Taking Im$%
[p^{0}]\gtrsim 0$ for the moment, we have the trajectories of fig. \ref%
{domdef2loops}. A\ trajectory lies above the $k_{x}$ integration path (which
may be deformed or not) if Re$[k_{x}]>p_{x}/2$ and below it if Re$%
[k_{x}]<p_{x}/2$. Since these conditions do not depend on the masses and $%
\mathbf{k}_{\perp }$, the trajectories lying on opposite sides of the $k_{x}$
integration path never intersect, so we do not need to worry about further
pinchings in the $k_{x}$ complex plane.

\begin{figure}[t]
\begin{center}
\includegraphics[width=9cm]{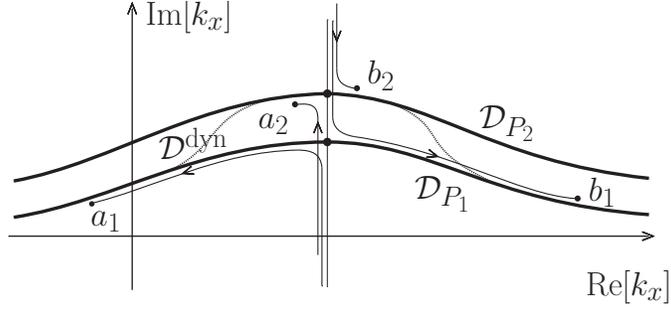}
\end{center}
\caption{\textquotedblleft Animation\textquotedblright\ of trajectories in
the $k_{x}$ complex plane in the presence of multiple thresholds with the
same external momentum $p$}
\label{squeezing}
\end{figure}

It may be helpful to see what happens with the help of a sort of animation.
Then we see that, say, the points $a_{1}$, $b_{1}$ lying on the trajectories
that approach the threshold $P_{1}$ arrive first, while the points $a_{2}$, $%
b_{2}$ lying on the trajectories that approach $P_{2}$ arrive later, as
shown in fig. \ref{squeezing}.

In figs. \ref{domdef2loops} and \ref{squeezing} the symbols $\mathcal{D}%
_{P_{i}}$, $i=1,2$, denote the $k_{x}$ integration domains that would
squeeze $\mathcal{\tilde{A}}_{P_{i}}$ onto the real axis if the threshold
were only $P_{i}$. In the presence of both thresholds, we deform the $k_{x}$
integration domain into a \textquotedblleft dynamic\textquotedblright\
domain $\mathcal{D}^{\text{dyn}}$ (i.e. a function of $p^{0}$) as follows.
At a first stage, when $a_{1}$, $b_{1}$ approach $P_{1}$ and $a_{2}$, $b_{2}$
are far away (in a neighborhood of the vertical Re$[k_{x}]=p_{x}/2$), $%
\mathcal{D}^{\text{dyn}}$ can be taken to be $\mathcal{D}_{P_{1}}$. At a
second stage, when $a_{1}$, $b_{1}$ are far away in a neighborhood of $%
\mathcal{D}_{P_{1}}$ and $a_{2}$, $b_{2}$ are approaching $P_{2}$, we
gradually deform $\mathcal{D}_{P_{1}}$ into $\mathcal{D}_{P_{2}}$, starting
from the vertical line towards the sides, as shown in fig. \ref{squeezing}.

What said about the trajectories displayed in figs. \ref{domdef2loops} and %
\ref{squeezing} can be repeated for the mirror trajectories obtained by
reflection with respect to the vertical line, which correspond to the case Im%
$[p^{0}]\lesssim 0$. Deforming the $k_{x}$ integration domain into $\mathcal{%
D}^{\text{dyn}}$ as explained, no pinching ever occurs in the complex $k_{x}$
plane as long as $|$Im$[p^{0}]|$ is sufficiently small and nonvanishing.
This means that the domain deformation can be finalized as expected, till
the region $\mathcal{\tilde{A}}_{P}$ squeezes completely onto the real axis
of the complex $p^{0}$ plane.

If the condition (\ref{pinchcond}) with $\mu \rightarrow \mu _{1}$, $%
M\rightarrow M_{1}$ is combined with the complex conjugate of the condition (%
\ref{pinchcond}) (where the conjugation does not act on the momenta) with $%
\mu \rightarrow \mu _{2}$, $M\rightarrow M_{2}$, then $P_{2}$ and the
trajectories approaching $P_{2}$ are reflected with respect to the real axis
and with respect to the vertical line Re$[k_{x}]=p_{x}/2$. The conclusions
reached above can easily be extended to this case.

It can also be seen that the branch points due to the square roots involved
in the expressions (\ref{pinchcond}), (\ref{pinchgen}) and (\ref{deno}) of $%
D_{\text{pinch}}$ are located away from the real axis of the $k_{x}$ complex
plane (if $\mu _{1}^{2}+\mathbf{k}_{\perp }^{2}+M_{1}^{2}$ and $\mu _{2}^{2}+%
\mathbf{k}_{\perp }^{2}+M_{2}^{2}$ are nonvanishing, which we may assume
here). Thus, if we choose $|p_{x}|$ large enough their branch cuts do not
intersect the trajectories and domains described so far.

\begin{figure}[t]
\begin{center}
\includegraphics[width=11cm]{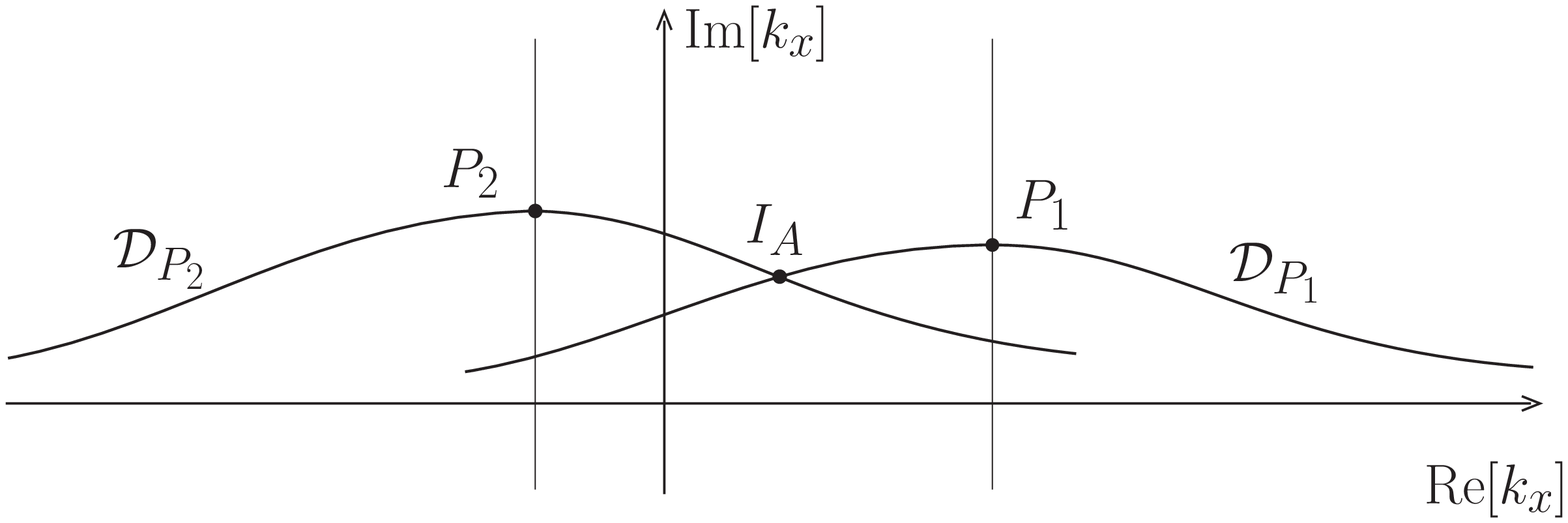}
\end{center}
\caption{Multiple thresholds with different external momenta $p$ and $q$}
\label{squeezing2}
\end{figure}

Now we consider the case of two LW thresholds $P_1$ and $P_2$ that depend on different
combinations $p$ and $q$ of external momenta, respectively. Again, we assume the worst
scenario for the loop momenta, which is when only one of them is involved.
This situation occurs, for example, in the triangle diagram. In $D=2$ we
have a picture such as the one of fig. \ref{squeezing2}.

We see that the two domains $\mathcal{D}_{P_{1}}$ and $\mathcal{D}_{P_{2}}$
may intersect in a point $I_{A}$, which is another true pinching. This kind
of pinching also occurs in ordinary models, where it gives the so-called 
\textit{anomalous threshold} \cite{anomalousthresh}. In two dimensions the
anomalous threshold of the triangle diagram is just a pole, but in higher
dimensions it is a branch point. Other intersections that may give anomalous
thresholds are those between $\mathcal{D}_{P_{1}}$ and the vertical line
crossing $P_{2}$, as well as the intersection between $\mathcal{D}_{P_{2}}$
and the vertical line crossing $P_{1}$.

Anomalous thresholds are known to appear in the diagrams that involve more
than one independent external momentum and have been studied at length in
the triangle and box diagrams. Basically, any time there are two external
momenta $p$ and $q$, or more, singularities of the form $\sim
1/f(p^{2},q^{2},p\cdot q)$ may appear, where $f(p^{2},q^{2},p\cdot q)$ is a
nontrivial function of the invariants that can be built with them. Anomalous
thresholds are associated with cuts that split the diagram in more than two
parts. It is known that they do not conflict with unitarity in ordinary
models. We will see that this property extends to the Lee-Wick models.
Ultimately, anomalous thresholds are sources of further complications, but
do not pose new conceptual challenges.

\begin{figure}[t]
\begin{center}
\includegraphics[width=11cm]{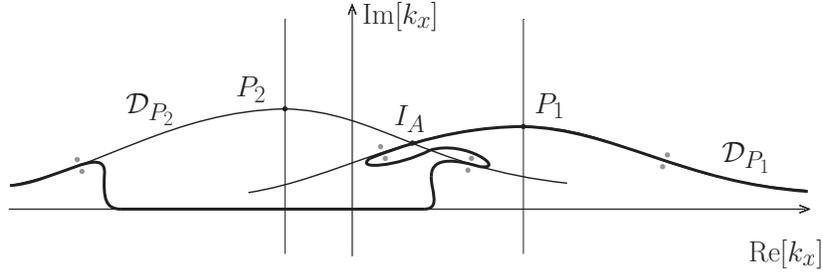}
\end{center}
\caption{Domain deformation in the presence of multiple thresholds with
different external momenta $p$ and $q$. The grey dots are the points $%
k_{x}^{\pm }(p)$ and $k_{x}^{\pm }(q)$}
\label{squeezing3}
\end{figure}

The dynamical squeezing can be achieved as follows. Consider the union $%
\mathcal{D}_{P_{1}}\cup \mathcal{D}_{P_{2}}$ and write it as $\mathcal{D}%
_{+}\cup \mathcal{D}_{-}$, where $\mathcal{D}_{+}$ (resp. $\mathcal{D}_{-}$)
is made of the superior (inferior) portions of $\mathcal{D}_{P_{1}}$ and $%
\mathcal{D}_{P_{2}}$ up to $I_{A}$. Start from the domain $\mathcal{D}_{+}$.
Consider the four trajectories $k_{x}^{\pm }(p)$ and $k_{x}^{\pm }(q)$ and
take energies $p^{0}$ and $q^{0}$ that make them stay in neighborhoods of $%
\mathcal{D}_{+}$. Let $p^{0}$ and $q^{0}$ grow till the trajectories
approach $I_{A}$. If the trajectories $k_{x}^{\pm }(q)$ arrive first and the
trajectories $k_{x}^{\pm }(p)$ arrive second, gradually deform $\mathcal{D}%
_{+}$ into the domain show in fig. \ref{squeezing3}. If $k_{x}^{\pm }(p)$
arrive first and $k_{x}^{\pm }(q)$ arrive second, take a domain deformation
that is symmetric to the one of fig. \ref{squeezing3} with respect to the
vertical line crossing $I_{A}$. The two possibilities correspond to the two
ways of circumventing the anomalous threshold $I_{A}$. When $p^{0}$ and $%
q^{0}$ grow more, it is enough to stretch the deformations just described.

The arguments given so far easily extend to $D>2$ and are exhaustive enough
to understand what happens in the most general case.

\section{Average continuation and difference continuation}

\setcounter{equation}{0}\label{avediffe}

When we start from the Euclidean version of the theory and perform the
nonanalytic Wick rotation, we must deform the integration domain on the loop
space momenta to overcome the LW thresholds. The domain deformation,
described in the previous section, is not easy to implement in general.
Fortunately, there is a shortcut to avoid it, which is the average
continuation.

In this section we formulate the average continuation and show that it
solves the nonanalytic Wick rotation and actually makes it unnecessary.
Precisely, the average continuation allows us to calculate the loop
integrals everywhere starting from the Euclidean region, or the region $%
\mathcal{\tilde{A}}_{0}$, without even entering the other regions $\mathcal{%
\tilde{A}}_{i}$, $i\neq 0$. We also study the difference continuation, which
is an elaboration of a rather familiar concept, but helps clarify the
properties of the average continuation by comparison.

The average continuation and the difference continuation are two noticeable
nonanalytic procedures to define a function of a complex variable $z$ beyond
a branch point $P$. The average continuation has to do with fakeons and
ultimately solves the Lee-Wick models. The difference continuation is at the
root of the cut diagrams. For simplicity, let us assume that $P$ is located
at the origin $z=0$. Let $f(z)$ denote the function we want to continue,
defined by choosing the branch cut to be the negative real axis.

\begin{figure}[t]
\begin{center}
\includegraphics[width=9cm]{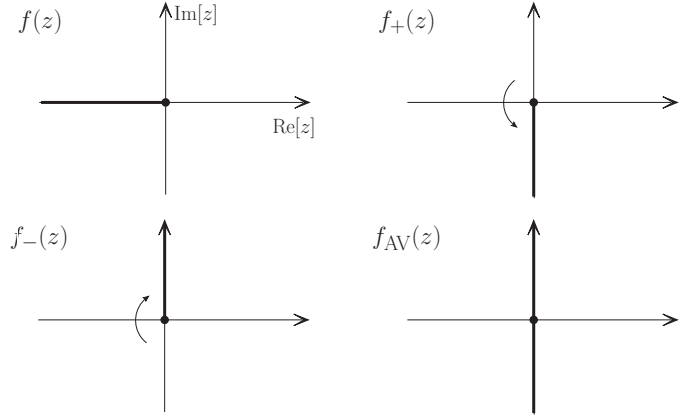}
\end{center}
\caption{Average continuation}
\label{avgcont}
\end{figure}

Referring to fig. \ref{avgcont}, define two other functions, $f_{+}(z)$ and $%
f_{-}(z)$, by choosing their branch cuts on the positive and negative
imaginary axes, respectively, i.e. $z=i\rho $ and $z=-i\rho $, with $\rho
\geqslant 0$. The average-continued function $f_{\text{AV}}(z)$ is defined
as the average of $f_{+}(z)$ and $f_{-}(z)$:%
\begin{equation}
f_{\text{AV}}(z)=\frac{1}{2}(f_{+}(z)+f_{-}(z)).  \label{favgcont}
\end{equation}

The imaginary axis divides the complex plane into two disjoint regions. This
means that $f_{\text{AV}}(z)$ is actually a collection of two analytic
functions: a \textit{superior function} $f_{>}(z)=\left. f_{\text{AV}%
}(z)\right\vert _{\mathrm{Re}[z]>0}$ and an \textit{inferior function} $%
f_{<}(z)\left. =f_{\text{AV}}(z)\right\vert _{\mathrm{Re}[z]<0}$.

The difference continuation is instead 
\begin{equation}
f_{\text{d}}(z)=\frac{1}{2}(f_{+}(z)-f_{-}(z)).  \label{fd}
\end{equation}%
Clearly, $f_{\text{d}}(z)=0$ in the half plane $\mathrm{Re}[z]>0$.

Among the properties of the average and difference continuations, we mention
that:

($i$) the inferior function $f_{<}(z)$ is uniquely determined by the
superior function $f_{>}(z)$, albeit in a\ nonanalytic way;

($ii$) the superior function $f_{>}(z)$ may or may not be determined by the
inferior function $f_{<}(z)$;

($iii$) the superior function cannot be analytically continued beyond $P$;

($iv$) it may or may not be possible to analytically continue the inferior
function beyond $P$.

($v$) if $g(z)$ is analytic or has a pole in $P$ and $h(z)\equiv f(z)g(z)$,
then $h_{\text{AV}}(z)=g(z)f_{\text{AV}}(z)$ and $f_{\text{d}}(z)=g(z)f_{%
\text{d}}(z)$;

($vi$) if $f(z)$ is real on the positive real axis, then $f_{\text{AV}}(z)$
and $f_{\text{d}}(z)$ are, respectively, real and purely imaginary on the
real axis.

In the case ($vi$), the value of $f_{\text{AV}}(z)$ on the negative real
axis is equal to the real part of either analytic continuation of $f(z)$ to
that half axis. Then we write 
\begin{equation}
f_{\text{AV}}(z)=\mathrm{Re}[f(z)],  \label{reim}
\end{equation}%
on the whole real axis.

\begin{figure}[t]
\begin{center}
\includegraphics[width=9cm]{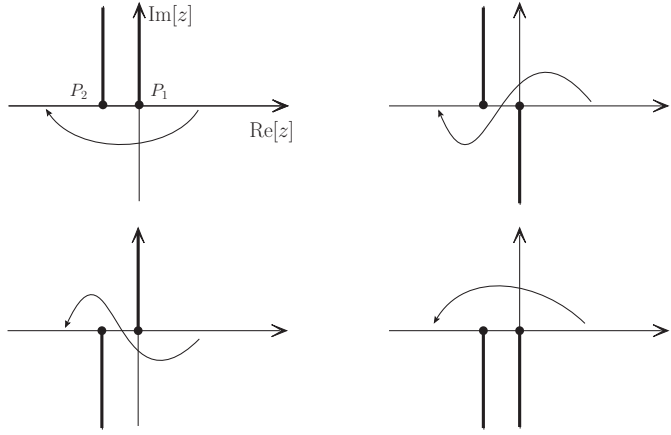}
\end{center}
\caption{Average continuation with more branch points}
\label{avgmo}
\end{figure}

More generally, if $f(z)$ has more distinct branch points on the real axis,
the average and difference continuations are defined by applying the rules
listed above to each branch point at a time. For example, let us study the
average continuation with two branch points $P_{1}$ and $P_{2}$. We have the
situation depicted in fig. \ref{avgmo}, which leads to three disjoint
regions: the half plane $\mathcal{A}_{1}=\{z:\mathrm{Re}[z]>\mathrm{Re}%
[P_{1}]\}$, the strip $\mathcal{A}_{2}=\{z:\mathrm{Re}[P_{2}]<\mathrm{Re}[z]<%
\mathrm{Re}[P_{1}]\}$ and the half plane $\mathcal{A}_{3}=\{z:\mathrm{Re}[z]<%
\mathrm{Re}[P_{2}]\}$. We come from $\mathcal{A}_{1}$, where the function is 
$f_{>}^{(1)}(z)\equiv f(z)$. We use the average continuation to overcome $%
P_{1}$ and reach the strip $\mathcal{A}_{2}$, which gives the inferior
function $f_{<}^{(1)}(z)$. Then we view $f_{<}^{(1)}(z)$ as the superior
function $f_{>}^{(2)}(z)$ for the second step and apply the average
continuation again, to overcome $P_{2}$ and go from $\mathcal{A}_{2}$ to $%
\mathcal{A}_{3}$. So doing, we obtain the new inferior function $%
f_{<}^{(2)}(z)$ for $\mathcal{A}_{3}$. At the end, $f_{\text{AV}}(z)$ is
equal to $f(z)$ in $\mathcal{A}_{1}$, $f_{<}^{(1)}(z)$ in $\mathcal{A}_{2}$
and $f_{<}^{(2)}(z)$ in $\mathcal{A}_{3}$.

When the branch points coincide, we must first deform them to make them
distinct (by varying the masses), then apply the procedure just described
and, finally, take the limit that makes them coincide. For example, consider
a diagram $G$ made of two diagrams $G_{1}$ and $G_{2}$ with one vertex in
common. The average-continued function $G_{\text{AV}}(z)$ associated with $G$
must clearly be the product $G_{1\text{AV}}(z)G_{2\text{AV}}(z)$ of the
average-continued functions $G_{1\text{AV}}(z)$ and $G_{2\text{AV}}(z)$
associated with $G_{1}$ and $G_{2}$. However, if $G_{1}$ and $G_{2}$ have
coinciding branch points, it may be tricky to satisfy this property.
Consider fig. \ref{avgmo} again: if $P_{1}=P_{2}$, we miss the paths shown
in the second and third figure, so we may obtain a wrong continuation. For
example, if $f(z)=\sqrt{z}$ is the superior function associated with $%
G_{1}=G_{2}$, then $f^{2}(z)=z$ is the superior function associated with $G$%
. However, $z$ has no branch point at all, rather than having two coinciding
branch points, so it cannot give the correct result. Instead, if we replace $%
f^{2}(z)$ with $\sqrt{z}\sqrt{z-a}$, with $a>0$, perform the average
continuation and let $a$ tend to zero at the end, we get the correct result.
The outcome is independent of the deformation. Indeed, if we exchange the
points $P_{1}$ and $P_{2}$ in fig. \ref{avgmo}, we simply exchange the
second and third figure, but in the limit $P_{2}\rightarrow P_{1}$ the
result does not change.

When a function $f(z_{1},\cdots ,z_{n})$ depends on $n>1$ complex variables
and there is a unique threshold, the singularities (solutions of $1/f=0$)
are generically a subspace $\mathcal{S}\subset \mathbb{C}^{n}$ of
codimension two and the branch subspaces $\mathcal{V}$ have codimension one,
with $\mathcal{S}=\partial \mathcal{V}$. Thus, there are still two ways to
analytically continue the function from $\mathbb{C}^{n}\backslash \mathcal{V}
$ beyond $\mathcal{S}$ to a neighborhood $\mathcal{A}$ of $\mathcal{V}$.
Again, the average continuation is half the sum of the two.

In the presence of several thresholds, we have several subspaces $\mathcal{V}
$. Their intersections give new regions $\mathcal{A}$. To reach the
intersection of two subspaces $\mathcal{V}$ we must perform two average
continuations in different variables. It is easy to check that the result is
independent of the order of the continuations. For example, let $n=2$ and $%
\mathcal{V}_{i}=\{(z_{1},z_{2}):\mathrm{Re}[z_{i}]>0\}$, $i=1,2$. Then we
can reach the intersection $\mathcal{V}_{1}\cap \mathcal{V}_{2}$ either by
first average-continuing in $z_{1}$ and then in $z_{2}$, or vice versa, but
the result does not change. The argument easily extends to multiple
intersections.

We define the average continuation recursively. Consider an arbitrary
diagram $G$. Deform the masses, so that the LW thresholds (\ref{thresh}) are
all distinct. Let $G_{i_{1}}$ denote the result of the average continuation
in some analytic region $\mathcal{A}_{i_{1}}$, already reached, with
nonvanishing widths $\epsilon $. In the zeroth step, we take the result $%
G_{0}$ of the loop integral in the main region $\mathcal{A}_{0}$. We want to
reach a new analytic region $\mathcal{A}_{i_{2}}$ above some LW threshold $P$%
. Redefine the external momenta $p_{1},\cdots ,p_{n}$ so that $P$ reads $%
p_{j}^{0}=\sqrt{\mathbf{p}_{j}^{2}+\tilde{M}^{2}}$ for some $p_{j}$ and some
combination $\tilde{M}$ of masses. Assume that an open-ball neighborhood $%
\mathcal{U}_{P}$ of $P$ belongs to $\mathcal{A}_{i_{1}}$, apart from the
points of the half line $\mathcal{O}_{P}$ with $\mathrm{Re}[\hat{p}%
_{j}^{0}]>0$, \textrm{Im}$[\hat{p}_{j}^{0}]=0$ in the $p_{j}^{0}$ complex
plane, where $\hat{p}_{j}^{0}\equiv p_{j}^{0}-\sqrt{\mathbf{p}_{j}^{2}+%
\tilde{M}^{2}}$. The average-continued function in $\mathcal{U}_{P}\cap 
\mathcal{O}_{P}$ is%
\begin{equation}
G_{\text{AV}}(p_{j}^{0})=\lim_{\delta \rightarrow 0^{+}}\frac{1}{2}\left[
G(p_{j}^{0}+i\delta )+G(p_{j}^{0}-i\delta )\right] .  \label{general}
\end{equation}%
Here and below, the dependence on $\mathbf{p}_{j}$ and the other external momenta is
understood. After the evaluation of $G_{\text{AV}}$, the deformed masses are
sent back to their original values and the result found in $\mathcal{U}%
_{P}\cap \mathcal{O}_{P}$ is extended to a neighborhood of $\mathcal{O}_{P}$
by analytic continuation, which defines the analytic region $\mathcal{A}%
_{i_{2}}$ above the threshold, as explained before. The operations (\ref%
{general}) must be applied to every LW threshold.

The relation between the average continuation and the nonanalytic Wick
rotation can be proved as follows. Let $\mathcal{\tilde{A}}_{i_{2}}$ denote
the region identified by the condition $D_{\text{pinch}}=0$, where $D_{\text{%
pinch}}$ is given by formula (\ref{pinchgen}). The behavior of the loop
integral around the pinching singularity inside $\mathcal{\tilde{A}}_{i_{2}}$
is, after integrating on the loop energies by means of the residue theorem, 
\begin{equation*}
\sim \int_{\mathcal{D}_{\mathbf{k},\mathbf{q}}}\frac{\prod\nolimits_{i=1}^{n}%
\mathrm{d}^{D-1}\mathbf{k}_{i}\prod\nolimits_{j=1}^{r-1}\mathrm{d}^{D-1}%
\mathbf{q}_{j}}{D_{\text{pinch}}(p_{j}^{0},\mathbf{k},\mathbf{q})},
\end{equation*}%
where $\mathcal{D}_{\mathbf{k},\mathbf{q}}$ is the integration domain on the
loop space momenta $\mathbf{k}$ and $\mathbf{q}$. The denominator is a
complex function, so its vanishing amounts to two real conditions. Write%
\begin{equation*}
D_{\text{pinch}}(p_{j}^{0},\mathbf{k},\mathbf{q})=-p_{j}^{0}+f_{j}(\mathbf{k}%
,\mathbf{q})-ig_{j}(\mathbf{k},\mathbf{q}),
\end{equation*}%
where $f_{j}$ and $g_{j}$ are real functions. When $\mathcal{D}_{\mathbf{k},%
\mathbf{q}}$ is deformed, the region $\mathcal{\tilde{A}}_{i_{2}}$ is
deformed as well. We have to arrange the domain deformation so as to squeeze 
$\mathcal{\tilde{A}}_{i_{2}}$ onto $\mathcal{O}_{P}$. Note that the deformed
integration domain may depend on the external momentum $p$, as discussed in
the previous section. We denote it by $\mathcal{D}_{\mathbf{k},\mathbf{q}}^{%
\text{def}}(p)$.

\begin{figure}[t]
\begin{center}
\includegraphics[width=10cm]{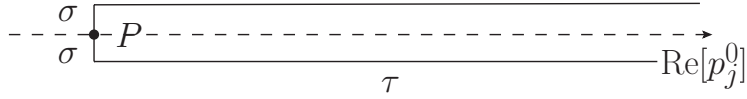}
\end{center}
\caption{Domain deformation}
\label{defo}
\end{figure}

Referring to fig. \ref{defo}, we arrange $\mathcal{D}_{\mathbf{k},\mathbf{q}%
}^{\text{def}}(p)$ so that $\mathcal{\tilde{A}}_{i_{2}}$ turns into half a
strip $\mathcal{\tilde{A}}_{i_{2}}^{\text{def}}$ of thickness $2\sigma $
centered in $\mathcal{O}_{P}$. The parameter $\sigma $ will later tend to
zero, to complete the domain deformation and squeeze $\mathcal{\tilde{A}}%
_{i_{2}}^{\text{def}}$ onto $\mathcal{O}_{P}$. When the loop momenta span
the domain $\mathcal{D}_{\mathbf{k},\mathbf{q}}^{\text{def}}(p)$, the
external momenta span the region $\mathcal{\tilde{A}}_{i_{2}}^{\text{def}}$.
We can use this map $\mathcal{D}_{\mathbf{k},\mathbf{q}}^{\text{def}%
}(p)\rightarrow \mathcal{\tilde{A}}_{i_{2}}^{\text{def}}$ to make a change
of variables such that $-\mathrm{Re}[p_{j}^{0}]+f_{j}(\mathbf{k},\mathbf{q}%
)=\tau $ and $\mathrm{Im}[p_{j}^{0}]+g_{j}(\mathbf{k},\mathbf{q})=\sigma
\eta $. Then, in spacetime dimensions $D$ greater than or equal to three,
the integral gets the form%
\begin{equation}
\int_{-\Delta }^{\infty }\mathrm{d}\tau \int_{-1}^{1}\mathrm{d}\eta \frac{%
h(\tau ,\sigma \eta )}{\tau -i\sigma \eta },  \label{inter}
\end{equation}%
where $\Delta >0$ and $h$ is regular at $\tau =\eta =0$. We understand that
the integral over the remaining variables has already been made. When $%
\sigma $ tends to zero, we obtain%
\begin{equation}
\int_{-\Delta }^{\infty }\mathrm{d}\tau \int_{-1}^{1}\mathrm{d}\eta \hspace{%
0.01in}h(\tau ,0)\left( \mathcal{P}\frac{1}{\tau }+i\pi \mathrm{sgn}(\eta
)\delta (\tau )\right) =2\int_{-\Delta }^{\infty }\mathrm{d}\tau \hspace{%
0.01in}h(\tau ,0)\mathcal{P}\frac{1}{\tau },  \label{resu}
\end{equation}%
where $\mathcal{P}$ denotes the principal value and $\mathrm{sgn}$ is the
sign function. This is the result of the nonanalytic Wick rotation. 

To perform the average continuation, we replace $p_{j}^{0}$ by $%
p_{j}^{0}+i\delta $, with $\delta $ real and small. Then, we first take $%
\sigma $ to zero keeping $|\delta |>0$ (which amounts to squeezing the
region $\mathcal{\tilde{A}}_{i_{2}}^{\text{def}}$ onto $\mathcal{O}_{P}$).
At a second stage, we study the limits $\delta \rightarrow 0^{+}$ and $%
\delta \rightarrow 0^{-}$. So doing, we approach $\mathcal{O}_{P}$ from
above ($\mathrm{Im}[\hat{p}_{j}^{0}]>0$, $\delta \rightarrow 0^{+}$) and
from below ($\mathrm{Im}[\hat{p}_{j}^{0}]<0$, $\delta \rightarrow 0^{-}$).
Since $|\delta |$ is small, the integral (\ref{inter}) becomes%
\begin{eqnarray}
\int_{-\Delta }^{\infty }\mathrm{d}\tau \int_{-1}^{1}\mathrm{d}\eta \frac{%
h(\tau ,\sigma \eta )}{\tau -i\delta -i\sigma \eta } &&\underset{\sigma
\rightarrow 0}{\longrightarrow }\int_{-\Delta }^{\infty }\mathrm{d}\tau
\int_{-1}^{1}\mathrm{d}\eta \frac{h(\tau ,0)}{\tau -i\delta }  \notag \\
&&\underset{\delta \rightarrow 0^{\pm }}{\longrightarrow }2\int_{-\Delta
}^{\infty }\mathrm{d}\tau \hspace{0.01in}h(\tau ,0)\left( \mathcal{P}\frac{1%
}{\tau }\pm i\pi \delta (\tau )\right) .  \label{resud}
\end{eqnarray}%
Averaging the two outcomes, we get (\ref{resu}) again. Thus, the nonanalytic
Wick rotation and the average continuation give the same results, as claimed.

With multiple thresholds the conclusions are the same, as long as the
threshold locations are distinct, as emphasized before. For two thresholds
located in $\tau _{1}$ and $\tau _{2}$, we have integrals of the form%
\begin{equation}
\int_{-\Delta }^{\infty }\mathrm{d}\tau \int_{-1}^{1}\mathrm{d}\eta \frac{%
h(\tau ,\sigma \eta )}{(\tau -\tau _{1}-i\delta _{1}-i\sigma \eta )(\tau
-\tau _{2}-i\delta _{2}-i\sigma \eta )}.  \label{2tres}
\end{equation}%
If $\tau _{1}\neq \tau _{2}$ the distributions of the form $\delta (\tau
-\tau _{1})\delta (\tau -\tau _{2})$ that would appear in the limits $\sigma
\rightarrow 0$, $\delta _{1}\rightarrow 0^{\pm }$, $\delta _{2}\rightarrow
0^{\pm }$, vanish. Note that they are multiplied by a power $\eta ^{2}$,
which is not killed by the integral over $\eta $. The distributions%
\begin{equation*}
\mathcal{P}\frac{1}{\tau -\tau _{1}}\delta (\tau -\tau _{2}),\qquad \mathcal{%
P}\frac{1}{\tau -\tau _{2}}\delta (\tau -\tau _{1}),
\end{equation*}%
are instead killed by the integral over $\eta $ (or the averages over\ the
limits $\delta _{1}\rightarrow 0^{\pm }$ and $\delta _{2}\rightarrow 0^{\pm
} $), so in the end we remain with%
\begin{equation*}
\mathcal{P}\frac{1}{\tau -\tau _{1}}\mathcal{P}\frac{1}{\tau -\tau _{2}},
\end{equation*}%
both in the case of the average continuation and in the case of the
nonanalytic Wick rotation.

The arguments and conclusions easily extend to $D=2$ once the integrals over 
$\eta $ that appear in formulas (\ref{inter}), (\ref{resu}), (\ref{resud})
and (\ref{2tres}) are replaced by sums over the values $\eta =-1$ and $\eta
=1$.

\begin{figure}[t]
\begin{center}
\includegraphics[width=8cm]{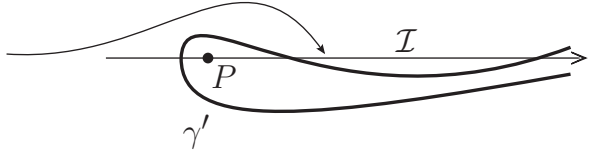}
\end{center}
\caption{Average continuation combined with the domain deformation}
\label{AVdeform}
\end{figure}

We conclude this section by mentioning other integral representations of the
average continuation, which will be useful for the proof of perturbative
unitarity. For definiteness, we take a unique LW threshold $P$ and assume
that it is located on the real axis. We deform the integration domain $%
\mathcal{D}_{\mathbf{k},\mathbf{q}}$ to a $\mathcal{D}_{\mathbf{k},\mathbf{q}%
}^{+\text{def}}$ such that the boundary curve $\gamma $ of fig. \ref%
{completo} is turned into a curve $\gamma ^{\prime }$ like the one of fig. %
\ref{AVdeform}. Then we consider the loop integral obtained by replacing the
domain $\mathcal{D}_{\mathbf{k},\mathbf{q}}$ with $\mathcal{D}_{\mathbf{k},%
\mathbf{q}}^{+\text{def}}$. Clearly, this integral representation allows us
to move analytically from the portion of the real axis that is located below
the intersection with $\gamma ^{\prime }$ to an interval $\mathcal{I}$ of
the real axis above $P$, without encountering LW pinchings. Let $\mathcal{J}%
_{+}$ denote the result of the loop integral calculated in $\mathcal{I}$
following this procedure. At a second stage, we make a mirror deformation $%
\mathcal{D}_{\mathbf{k},\mathbf{q}}^{-\text{def}}$, so as to obtain a
picture where $\gamma $ is turned into the reflection of $\gamma ^{\prime }$
with respect to the real axis. We calculate the loop integral in $\mathcal{I}
$ and call the result $\mathcal{J}_{-}$. The integral representation of the
average continuation in $\mathcal{I}$ is $(\mathcal{J}_{+}+\mathcal{J}%
_{-})/2 $. We can further deform the domains $\mathcal{D}_{\mathbf{k},%
\mathbf{q}}^{\pm \text{def}}$ so as to stretch $\mathcal{I}$ to the whole $%
\mathcal{O}_{P}$. The construction easily generalizes to LW thresholds that
are not on the real axis and to multiple LW thresholds. 

\section{Average continuation in various dimensions}

\setcounter{equation}{0}

\label{avedime}

In this section we illustrate the average continuation in examples related
to typical loop integrals.

The first example is $f(z)=\ln z$, with the branch cut on the negative real
axis. The functions $f_{\pm }(z)$ of the previous section are $\ln (z\pm
i\epsilon )$, so, by formula (\ref{favgcont}), the average-continued
function turns out to be 
\begin{equation}
f_{\text{AV}}(z)=\frac{1}{2}\ln z^{2}.  \label{4dfake}
\end{equation}%
The imaginary axis divides the complex plane in two disjoint regions: the
half plane $\mathrm{Re}[z]>0$ and the half plane $\mathrm{Re}[z]<0$. The
superior function can be determined from the inferior function, but neither
of the two can be analytically continued beyond $z=0$. By comparison, the
Feynman prescription gives $\ln (z-i\epsilon )$.

The difference continuation gives%
\begin{equation}
f_{\text{d}}(z)=\left\{ 
\begin{tabular}{l}
$0$ for $\mathrm{Re}[z]>0,$ \\ 
$i\pi $ for $\mathrm{Re}[z]<0,$%
\end{tabular}%
\right.  \label{4dim}
\end{equation}%
which may be written as $i\pi \theta (-z)$, where $\theta (z)$ is the
complex $\theta $ function, equal to 1 for $\mathrm{Re}[z]>0$ and 0 for $%
\mathrm{Re}[z]<0$.

Note that the function $\ln z$ with $z\rightarrow -p^{2}$ is the value of
the bubble diagram of a massless scalar field in four dimensions, apart from
an overall factor and an additional constant. The Feynman prescription leads
to $\ln (-p^{2}-i\epsilon )$, while the average continuation leads to $f_{%
\text{AV}}(-p^{2})=(1/2)\ln (p^{2})^{2}$ \cite{LWgravmio}. If we squeeze the
half plane $\mathrm{Re}[z]<0$ onto the negative real axis, formula (\ref%
{4dim}) encodes the discontinuity of the amplitude of the bubble diagram,
i.e. the sum of the two cut diagrams associated with it, which is
proportional to\ $f_{\text{d}}(-p^{2})=i\pi \theta (p^{2})$.

As a second example, consider the function $f(z)=\sqrt{z}$. We find%
\begin{equation}
f_{\text{AV}}(z)=\left\{ 
\begin{tabular}{l}
$\sqrt{z}$ for $\mathrm{Re}[z]>0,$ \\ 
$0$ for $\mathrm{Re}[z]<0,$%
\end{tabular}%
\right. \qquad f_{\text{d}}(z)=\left\{ 
\begin{tabular}{l}
$0$ for $\mathrm{Re}[z]>0,$ \\ 
$i\sqrt{-z}$ for $\mathrm{Re}[z]<0.$%
\end{tabular}%
\right.  \label{avroot}
\end{equation}%
Here, the superior function cannot be determined from the inferior one,
which vanishes. The inferior function can be trivially continued beyond $z=0$%
, while the superior function obviously cannot.

In three dimensions, the bubble diagram of a massless scalar does not give a
logarithm, but $1/\sqrt{-p^{2}}$. The Feynman prescription leads to $1/\sqrt{%
-p^{2}-i\epsilon }$. If we use property ($v$) of section (\ref{avediffe})
with $g(z)=1/z$, $f(z)=\sqrt{z}$ and $h(z)=1/\sqrt{z}$, we find $f_{\text{AV}%
}(z)=0$ for $\mathrm{Re}[z]<0$. Again, the difference continuation is
proportional to the discontinuity of the bubble diagram.

\subsection{Four dimensions}

\begin{figure}[t]
\begin{center}
\includegraphics[width=6cm]{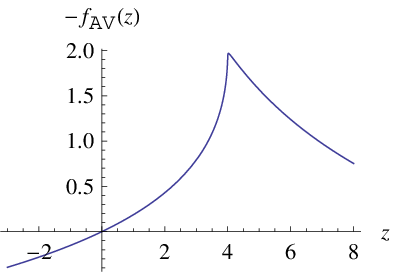}\qquad%
\includegraphics[width=6cm]{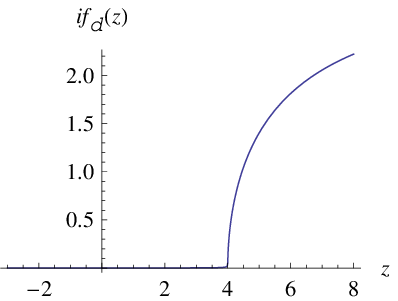}
\end{center}
\caption{Average and difference continuations of the massive fakeon}
\label{logAV}
\end{figure}

In the massive case, the bubble diagram of the standard scalar field in four
dimensions leads to the well-known expression%
\begin{equation*}
\int_{0}^{1}\mathrm{d}x\ln \left[ -p^{2}x(1-x)+m^{2}-i\epsilon \right] ,
\end{equation*}%
after renormalizing the divergent part. This function has branch cuts in $%
p^{2}=(2m)^{2}$. Switching to the dimensionless variable $z=p^2/m^2$, we are lead to study
the function%
\begin{equation*}
f(z)=\int_{0}^{1}\mathrm{d}x\ln \left[ 1-zx(1-x)\right] ,
\end{equation*}%
whose average continuation is straightforward, by formula (\ref{4dfake}),
and gives%
\begin{equation}
f_{\text{AV}}(z)=\frac{1}{2}\int_{0}^{1}\mathrm{d}x\ln \left[ (1-zx(1-x))^{2}%
\right] .  \label{mfake}
\end{equation}%
In fig. \ref{logAV} we show the plot of this function for $z$ real, together
with the plot of the difference continuation. 
The first plot has the typical form of the LW\ amplitudes around the LW
pinching \cite{LWformulation}. Basically, the average continuation turns the
ordinary scalar field into a massive fakeon (see the next section for
details), i.e. the massive version of the fake degree of freedom of ref. 
\cite{LWgravmio}.

Now, consider the LW propagator%
\begin{equation}
\frac{1}{2}\left( \frac{1}{p^{2}-iM^{2}}+\frac{1}{p^{2}+iM^{2}}\right) .
\label{lwbas}
\end{equation}%
The bubble diagram built with it has the LW threshold $p^{2}=2M^{2}$. Again,
in the Euclidean region $|\mathrm{Re}[p^{0}]|<|\mathbf{p}|$ we can evaluate
the loop integral straightforwardly by means of the Feynman parameters. We
have the sum of four contributions%
\begin{equation*}
r_{ab}(p^{2}/M^{2})=\frac{1}{4}\int \frac{\mathrm{d}^{D}k}{(2\pi )^{D}}\frac{%
1}{k^{2}-iaM^{2}}\frac{1}{(p-k)^{2}-ibM^{2}},
\end{equation*}%
where $a,b=+$ or $-$. The functions $r_{++}$ and $r_{--}$ can be
analytically continued to the whole real axis, because they are not
interested by LW pinchings. Renormalizing away the divergent part, their sum
is equal to $-ig(p^{2}/M^{2})/(8\pi )^{2}$, where%
\begin{equation}
g(t)\equiv \int_{0}^{1}\mathrm{d}x\ln H(x,t),\qquad H(x,t)\equiv
1+t^{2}x^{2}(1-x)^{2}.  \label{ppmm}
\end{equation}

For $p^{2}<0$, the sum of $r_{+-}$ and $r_{-+}$ is $-if_{0}(p^{2}/M^{2})/(8%
\pi )^{2}$, where%
\begin{equation}
f_{0}(t)\equiv \int_{0}^{1}\mathrm{d}x\ln K(x,t),\qquad K(x,t)\equiv
(1-2x)^{2}+t^{2}x^{2}(1-x)^{2}.  \label{rpm}
\end{equation}%
The function $f_{0}(t)$ does not give the correct result for $t>0$. Indeed,
it is symmetric under $t\rightarrow -t$ and not analytic in $t=0$ (which is
not a LW\ threshold). We have to analytically continue $f_{0}(t)$ from $t<0$
up to the LW threshold $t=2$. Then we have to average-continue it beyond the
LW threshold.

Observe that $K(x,t)$ has four zeros in $x$, which are $x=u(t)$, $x=u^{\ast
}(t)$, $x=$ $v(t)$ and $x=v^{\ast }(t)$, where%
\begin{equation*}
u(t)=\frac{1}{2}-\frac{i}{t}+\frac{i}{2t}\sqrt{4-t^{2}},\qquad v(t)=\frac{1}{%
2}-\frac{i}{t}-\frac{i}{2t}\sqrt{4-t^{2}}.
\end{equation*}%
We have to concentrate on the interval $0<t<2$. We see that $\mathrm{Im}%
[v(t)]$ does not vanish, while $\mathrm{Im}[u(t)]$ vanishes for $t=0$ and
only there. In that point, $u$ is equal to $1/2$, which belongs to the
integration path $0<x<1$. When $t$ grows and crosses the value $0$, two
zeros, $u(t)$ and $u^{\ast }(t)$, cross the integration path, while the
other two remain far away.

\begin{figure}[t]
\begin{center}
\includegraphics[width=6cm]{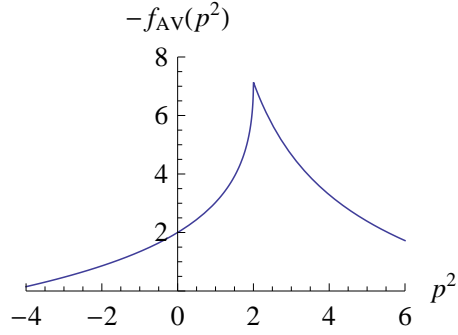}
\end{center}
\caption{LW fakeon}
\label{LWfakeon}
\end{figure}

It is simple to analytically continue the derivative $f_{0}^{\prime }(t)$
beyond $t=0$, because its integrand is meromorphic. We just have to add the
residues of the poles that cross the integration path, which are equal to $%
-2\pi iu^{\prime }(t)$ and $2\pi iu^{\prime \ast }(t)$. When we go back to
the primitive, we obtain, on the real axis, the function%
\begin{equation*}
f(t)=f_{0}(t)+\frac{2\pi }{t}\theta (t)\theta (2-t)\left( \sqrt{4-t^{2}}%
-2\right) ,
\end{equation*}%
which is indeed analytic for $t<2$.

At this point, it is easy to perform the average continuation above the LW
threshold $t=2$. Observe that the average continuations of $f_{0}(t)$ and $%
1/t$ are trivial, while the average continuation of the square root is zero,
by formula (\ref{avroot}). Thus, above the LW threshold we just have to drop
the square root. The final result is (on the real axis)%
\begin{equation}
f_{\text{AV}}(t)=f_{0}(t)+\frac{2\pi }{t}\theta (t)\theta (2-t)\left( \sqrt{%
4-t^{2}}-2\right) -\frac{4\pi }{t}\theta (t-2).  \label{fav}
\end{equation}%
Its plot is shown in fig. \ref{LWfakeon} and is very similar to the one of
the massive fakeon shown in the left picture of fig. \ref{logAV}. We can
call it \textit{Lee-Wick fakeon}. 

Repeating the arguments for the more general LW propagator%
\begin{equation}
\frac{1}{2}\left( \frac{1}{p^{2}-\mu ^{2}-iM^{2}}+\frac{1}{p^{2}-\mu
^{2}+iM^{2}}\right) ,  \label{mum}
\end{equation}%
and focusing on $r_{+-}=r_{-+}$ ($r_{++}$ and $r_{--}$ still being analytic
on the real axis) we get%
\begin{eqnarray*}
f_{0}(t,r) &=&\int_{0}^{1}\mathrm{d}x\ln \left[ (1-2x)^{2}+(r-tx(1-x))^{2}%
\right] , \\
f(t,r) &=&f_{0}(t,r)+\frac{2\pi }{t}\theta (t-4r)\theta (\sigma -t)(\sqrt{%
4+4rt-t^{2}}-2), \\
f_{\text{AV}}(t,r) &=&f(t,r)-\frac{4\pi }{t}\theta (t-\sigma ),
\end{eqnarray*}%
where $r=\mu ^{2}/M^{2}$ and $\sigma =2\sqrt{r^{2}+1}+2r$.

Finally, we study the nonanalytic Wick rotation of the Euclidean theory and
compare it to the average continuation. We work with the propagator (\ref%
{lwbas}). The average continuation of the amplitude on the real axis is%
\begin{equation}
\mathcal{M}_{\text{AV}}(p^{2},M^{2})=-\frac{1}{2(8\pi )^{2}}\left[ f_{\text{%
AV}}(p^{2}/M^{2})+g(p^{2}/M^{2})\right] ,  \label{mav}
\end{equation}%
where the combinatorial factor 1/2 is included.

\begin{figure}[t]
\begin{center}
\includegraphics[width=6cm]{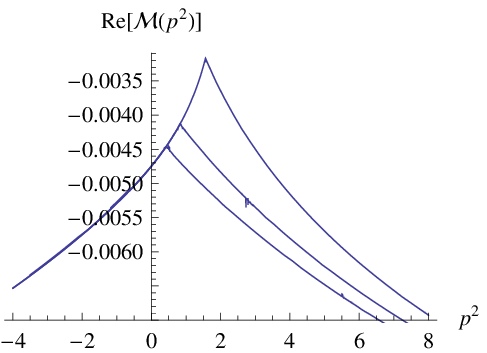}\qquad %
\includegraphics[width=6cm]{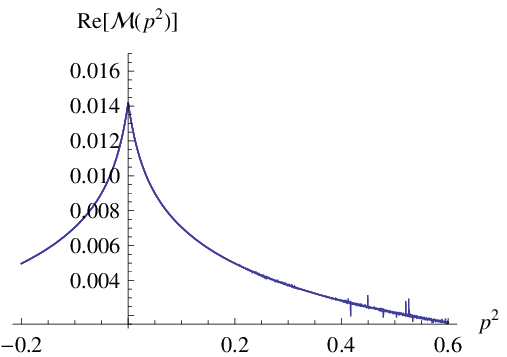}
\end{center}
\caption{Four-dimensional LW fakeon: numerical results from the nonanalytic
Wick rotation (with no domain deformation) for $|\mathbf{p}|=1,2,3$, with $%
M=1$ (left picture) and $M=1/20$ (right picture).}
\label{4Dfakeons}
\end{figure}

If we want to evaluate the amplitude by means of the nonanalytic Wick
rotation, we have to make the calculation inside the region $\mathcal{\tilde{%
A}}_{P}$ of fig. \ref{completo} and deform the integration domain on the
loop space momentum as explained in section \ref{thedomdef}, till $\mathcal{%
\tilde{A}}_{P}$ squeezes onto $\mathcal{O}_{P}$, which is the portion of the
real axis from $P$ to $+\infty $. The procedure is involved, but there
are situations where the region $\mathcal{\tilde{A}}_{P}$ is sufficiently
thin to make the actual deformation unnecessary. One such case is when the
LW scale $M$ is small. It does not even need to be so small, since in most
formulas it is raised to the forth power.

A measure of the violations of analyticity and Lorentz invariance, which
occur before the domain deformation, is given by the \textquotedblleft
distance\textquotedblright\ between the point $P$ and the point $P^{\prime }$
of fig. \ref{completo}, i.e. the difference between the values of $p^{2}$ in
such two points. Expanding the difference for $M$ small, we find%
\begin{equation}
\sim 2M^{2}-\frac{4M^{4}}{\mathbf{p}^{2}}.  \label{estim}
\end{equation}%
The first term is Lorentz invariant, so it controls the violation of
analyticity. The second term controls the Lorentz violation. We see that the
Lorentz violation is much smaller than the violation of analyticity.
Numerically, we should see an evident Lorentz violation for, say $|\mathbf{p}%
|=1$, $2$, $3$, $M=1$, and an approximately Lorentz invariant result already
for $M=1/20$, with the same values of $|\mathbf{p}|$. The two situations are
shown in fig. \ref{4Dfakeons}, which confirms what we have just said. From
left to right, the three plots are $|\mathbf{p}|=3$, $2$ and $1$. In the
first picture, where $M=1$, the plots superpose below the minimum $P^{\prime
}$, but evidently deviate from one another above $P^{\prime }$ and $P$. In
the second picture, which has $M=1/20$, the agreement is good everywhere.

\begin{figure}[t]
\begin{center}
\includegraphics[width=6cm]{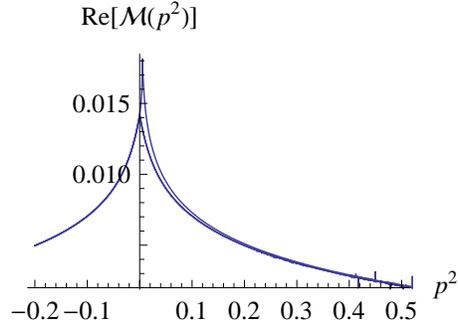}
\end{center}
\caption{Comparison between the numerical results from the nonanalytic Wick
rotation (with no domain deformation) and the result of the average
continuation for $M=1/20$.}
\label{compa4D}
\end{figure}

In fig. \ref{compa4D} we include the prediction of the average continuation
for $M=1/20$, which is the top graph. As predicted by the first term on the
right-hand side of formula (\ref{estim}), we see a discrepancy in the
interval $0<p^{2}\lesssim 2M^{2}\sim .005$ (caused by the missing domain
deformation) and agreement everywhere else. 

\subsection{Three dimensions}

In three dimensions the bubble diagram built with the propagator (\ref{lwbas}%
) gives the functions%
\begin{eqnarray*}
f_{0}(t)&=&\int_{0}^{1}\mathrm{d}x\left( \frac{1}{\sqrt{-tx(1-x)+i(1-2x)}}+%
\frac{1}{\sqrt{-tx(1-x)-i(1-2x)}}\right), \\
g(t)&=&\int_{0}^{1}\mathrm{d}x\left( \frac{1}{\sqrt{-tx(1-x)+i}}+\frac{1}{%
\sqrt{-tx(1-x)-i}}\right) .
\end{eqnarray*}%
Here it is more tricky to work with the integrands, so it is better to
eliminate the Feynman parameters by evaluating the integrals explicitly. In
the Euclidean region $t<0$ we find%
\begin{equation*}
f_{0}(t) =\frac{2i}{\sqrt{-t}}\ln \left( \frac{\sqrt{2}-i\sqrt{-t}}{\sqrt{2%
}+i\sqrt{-t}}\right) , \,\quad
g(t) =\frac{i}{\sqrt{-t}}\left[ \ln \left( \frac{2\sqrt{i}+\sqrt{-t}}{2%
\sqrt{i}-\sqrt{-t}}\right) +\ln \left( \frac{2\sqrt{-i}-\sqrt{-t}}{2\sqrt{-i}%
+\sqrt{-t}}\right) \right] .
\end{equation*}%
It is important to take such functions exactly as they are written, because
manipulations that look innocuous may actually conflict with the
determinations of the square roots and the logarithms. We have chosen to
write the formulas so that they have the correct expansions for $t\sim
-\infty $.

By formula (\ref{reim}), the average continuation on the real axis is just
the real part, which gives the bubble amplitude 
\begin{equation}
\mathcal{M}_{\text{AV}}(p^{2},M^{2})=\frac{1}{64\pi M}\mathrm{Re}%
[f_{0}(p^{2}/M^{2})+g(p^{2}/M^{2})].  \label{avdef}
\end{equation}

\begin{figure}[t]
\begin{center}
\includegraphics[width=6cm]{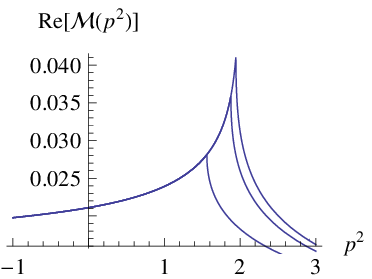}\qquad %
\includegraphics[width=6cm]{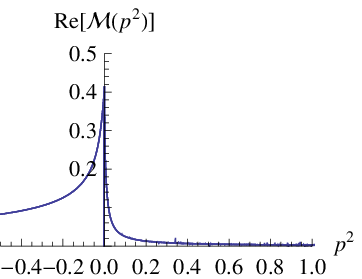}
\end{center}
\caption{Three-dimensional LW fakeon: numerical results from the nonanalytic
Wick rotation with no domain deformation for $|\mathbf{p}|=1,1/2,1/3$, $M=1$
(left picture) and $|\mathbf{p}|=1,2,3$, $M=1/20$ (right picture)}
\label{3DFakeons}
\end{figure}

As in four dimensions, the nonanalytic Wick rotation exhibits, before the
domain deformation, violations of analyticity and Lorentz invariance. They
are apparent at $M=1$, and, say, $|\mathbf{p}|=1,1/2,1/3$, as confirmed by
the left picture of fig. \ref{3DFakeons}. By the estimate (\ref{estim}), we
expect that Lorentz invariance is quickly recovered at, say, $M=1/20$, which
is confirmed by the right picture of fig. \ref{3DFakeons}, where $|\mathbf{p}%
|=1,2,3$. Zooming in, it is possible to observe a slight discrepancy around $%
p^{2}=0$, which is the violation of analyticity due to missing domain
deformation and estimated by the first term of (\ref{estim}). 

\begin{figure}[t]
\begin{center}
\includegraphics[width=6cm]{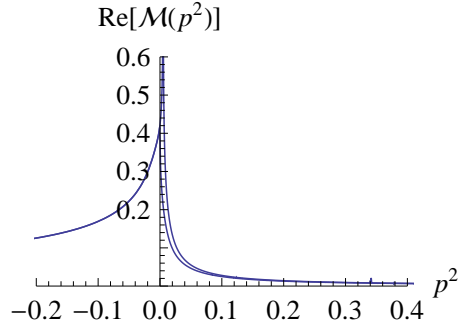}
\end{center}
\caption{Three-dimensional LW fakeon: comparison between the average
continuation and the nonanalytic Wick rotation with no domain deformation at 
$M=1/20$}
\label{compa3D}
\end{figure}

By applying formula (\ref{avdef}), we can compare the results for $M=1/20$
with the ones of the average continuation. This gives fig. \ref{compa3D}.
Again, we see small discrepancies between $P^{\prime }$ and $P$, due to the
missing domain deformation, but agreement below $P^{\prime }$, where no
domain deformation is required, and above $P$, where the effects of the
domain deformation are negligible.

\subsection{Two dimensions}

In two dimensions the bubble diagram with propagators (\ref{lwbas}) gives,
in the Euclidean region $t<0$, a result proportional to the sum $%
f_{0}(t)+g(t)$, where 
\begin{equation*}
f_{0}(t)=-2t\int_{0}^{1}\mathrm{d}x\hspace{0.01in}\hspace{0.01in}\frac{x(1-x)%
}{K(x,t)},\qquad g(t)=-2t\int_{0}^{1}\mathrm{d}x\hspace{0.01in}\hspace{0.01in%
}\frac{x(1-x)}{H(x,t)}.
\end{equation*}%
As before, the integrand of $f_{0}$ has four singularities on the imaginary
axis of the complex $x$ plane. Two of them cross the $x$ integration path
when $t$ varies from negative to positive values, while the other two do not
cross the integration path. Since the singularities are poles, the
difference $f(t)-f_{0}(t)$ for $0<t<2$ can be easily calculated by summing
the two residues, multiplied by $2\pi i$. We find%
\begin{equation*}
f(t)=f_{0}(t)+\frac{4\pi }{\sqrt{4-t^{2}}}\theta (t)\theta (2-t).
\end{equation*}%
Then, $f_{\text{AV}}(t)=f(t)$ on the whole real axis. Indeed, we know that
the average continuation of the function $\sim 1/\sqrt{t}$ is zero below $%
t=0 $.

From the point of view of the nonanalytic Wick rotation, the two-dimensional
models are a bit different from the models in dimensions greater than or
equal to three. The reason is that in two dimensions the LW pinching occurs
only at the boundary of the region $\mathcal{\tilde{A}}_{P}$ of fig. \ref%
{completo}, but not inside. The result of a loop integral in $\mathcal{O}%
_{P} $ is Lorentz invariant and analytic even before making the domain
deformation. The only Lorentz violation we find in the intermediate steps is
due to the fact that $\mathcal{\tilde{A}}_{P}$ extends to $P^{\prime }$. To
recover Lorentz invariance, it is sufficient to ignore the function found
inside $\mathcal{\tilde{A}}_{P}$ below $P$ and analytically extend the
function found in $\mathcal{\tilde{A}}_{0}$ from $P^{\prime }$ to $P$.

\begin{figure}[t]
\begin{center}
\includegraphics[width=8cm]{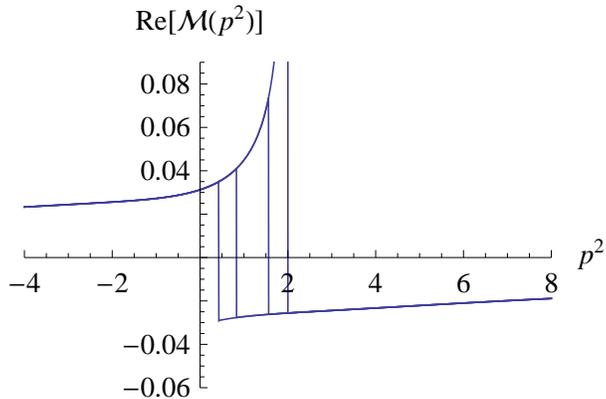}
\end{center}
\caption{Two-dimensions: numerical results from the nonanalytic Wick
rotation with no domain deformation and comparison with the average
continuation}
\label{compa}
\end{figure}

We can show these facts numerically, by plotting the results of the
calculations for real $p^{0}$ around the points $P$, $P^{\prime }$, with
various values of $|\mathbf{p}|$. In fig. \ref{compa} we see four vertical
lines. The first three, from the left to the right, correspond to $|\mathbf{p%
}|=3,2,1$, with $M=1$. Their locations are those of the point $P^{\prime }$.
We see that each pair of plots agree both below the smaller $P^{\prime }$
and above the larger $P^{\prime }$.

The forth vertical line of fig. \ref{compa} corresponds to the result of the
average continuation. We see that the nonanalytic Wick rotation with no
domain deformation and the average continuation agree both below $P^{\prime
} $ and above $P$, even if $M$ is not small with respect to $|\mathbf{p}|$.

In conclusion, a great simplification occurs in two dimensions, where the
domain deformation is not strictly required to make calculations by means of
the nonanalytic Wick rotation. At the same time, we have learned how
powerful the average continuation is, because it drastically reduces the
calculational effort in all dimensions.

\section{Fakeons}

\setcounter{equation}{0}\label{fakeons}

We have seen that the average continuation is a simple operation to overcome
branch points. Then, it is natural to inquire what happens if we apply it to
a physical degree of freedom. Consider for example, the bubble diagram of
ordinary scalar fields, which can be formally obtained by letting $M$ tend
to infinity in formula (\ref{bub}). The propagator just has the circled
poles of fig. \ref{prop}. After taking $\epsilon \rightarrow 0$, the bubble
loop integral has two branch points on the real axis at $%
p^{2}=(m_{1}+m_{2})^{2}$. The branch cuts are the half lines $p^{2}\geqslant
(m_{1}+m_{2})^{2}$ on the real axis. An $\epsilon $ different from zero
gives the familiar Feynman prescription, which displaces the branch cuts a
little bit from the real axis and thereby allows us to define the loop
integrals above the thresholds by analytic continuation from the segment $%
p^{2}<(m_{1}+m_{2})^{2}$ to the half lines $p^{2}\geqslant (m_{1}+m_{2})^{2}$%
. The displacements in the bubble diagram and its conjugate diagram are
symmetric with respect to the real axis. This originates the discontinuity
of the amplitude and, ultimately, the propagating degree of freedom. After
subtracting the ultraviolet divergence, the diagram gives, in the massless
case $m_{1}=m_{2}=0$,%
\begin{equation}
-\frac{i}{2(4\pi )^{2}}\ln \frac{-p^{2}-i\epsilon }{\mu ^{2}},
\label{amplaf}
\end{equation}%
where we have included the combinatorial factor $1/2$.

The average continuation can be viewed as an alternative prescription to
define the loop integral above the thresholds. If we forget about $\epsilon $%
, by setting it to zero from the start, we can still define the amplitude
unambiguously above the thresholds by means of formula (\ref{4dfake}), in
which case the result becomes (for $p$ real)%
\begin{equation}
-\frac{i}{4(4\pi )^{2}}\ln \frac{(p^{2})^{2}}{\mu ^{4}}.  \label{ampla}
\end{equation}%
The discontinuity is absent, so we have no propagating degree of freedom.
Equivalently, we can say that we have a fakeon, a fake degree of freedom.
The average continuation makes the physical field disappear from the
spectrum.

At the level of the Feynman rules, the fakeon prescription can be formulated
as follows. We replace the propagator $1/(p^{2}-m^{2})$ with \cite{LWgravmio}
\begin{equation}
\frac{p^{2}-m^{2}}{(p^{2}-m^{2})^{2}+\mathcal{E}^{4}},  \label{newdistr}
\end{equation}%
which coincides with (\ref{mum}) apart from the notation, and let $\mathcal{E%
}$ tend to zero at the very end. The limit $\mathcal{E}\rightarrow 0$ is
regular, since it is just a prescription for the propagator.

The results of this paper apply to the theories whose elementary fields have
free propagators that contain:

($i$) ordinary poles, treated by means of the Feynman prescription (with
infinitesimal widths $\epsilon $);

($ii$) LW poles, with finite LW scales $M$;

($iii$) fakeons, defined by means of the prescription (\ref{newdistr}), with
infinitesimal LW scales $\mathcal{E}$.

The widths $\epsilon $ must tend to zero first and the LW scales $\mathcal{E}
$ must tend to zero last. At $\mathcal{E}>0$ we have a LW model, because the
poles of type ($iii$) are just like the LW poles of type ($ii$). In that
case, we make the computations by means of the nonanalytic Wick rotation or
the average continuation. The results of the next section ensure that the
theory is perturbatively unitary for $\epsilon \rightarrow 0$ at $\mathcal{E}%
>0$. If we let $\mathcal{E}$ tend to zero at the very end, perturbative
unitarity is preseved, since it holds for every nonzero $\mathcal{E}$.

We can retrieve the fakeon (\ref{newdistr}) from the results of the previous
section, by taking the limit $M\rightarrow 0$. For example, if we let $M$
tend to zero in formula (\ref{mav}), we get $-i$ times (\ref{ampla}), which
is correct, since for $M\rightarrow 0$ the propagator (\ref{lwbas}) is the
usual scalar propagator $1/p^{2}$ endowed with the fakeon prescription (\ref%
{newdistr}). In three dimensions we can take the limit $M\rightarrow 0$ of
formula (\ref{avdef}), which gives $\theta (-p^{2})/(16\sqrt{-p^{2}})$.

While the LW degrees of freedom ($ii$) require higher derivatives and have
finite LW scales $M$, the fake degrees of freedom ($iii$) can be introduced
even without higher derivatives and have infinitesimal LW\ scales $\mathcal{E%
}\rightarrow 0$. Yet, there is not a deep difference between the two. In this
respect, recall that the numerators of the propagators, such as the one of (%
\ref{newdistr}), are not important in the study of the LW pinchings. From
now on, we call fakeons both the LW degrees of freedom ($ii$) and the fake
degrees of freedom ($iii$). We may speak of fakeon thresholds, instead of LW
thresholds, fakeon scales, and so on. We call fakeon theories the theories
that involve fakeons (of LW type or not)\ besides ordinary physical degrees
of freedom. Every result of this paper applies to the most general fakeon
theory in dimensions $D$ greater than or equal to 2.

Observe that if we plan to take $M$, or $\mathcal{E}$, to zero, the
nonanalytic Wick rotation simplifies enormously, because there is no need to
make the domain deformation. A quick way to see this is provided by formula (%
\ref{estim}), which gives an estimate of the analyticity violations and the
Lorentz violations that occur prior to the domain deformation. Clearly, they
both disappear in the limit $M\rightarrow 0$. A more detailed argument can
be provided by means of formula (\ref{deno}). Assume that we may have a LW
pinching, i.e. $n=n_{+}+n_{-}>0$. The pinching condition $D_{\text{pinch}}=0$%
, which defines the regions $\mathcal{\tilde{A}}_{i}$, $i\neq 0$, implies 
\begin{equation*}
|\mathrm{Im}[p^{0}]|\leqslant \sum_{i=1}^{n_{+}}|\mathrm{Im}[\Omega ^{+}(%
\mathbf{k}_{i})]|+\sum_{i=n_{+}+1}^{n}|\mathrm{Im}[\Omega ^{-}(\mathbf{k}_{i}%
\mathbf{)}]|=\sum_{i=1}^{n}\eta _{-}(\mathbf{k}_{i}^{2}+\mu ^{2}\mathbf{)}%
\leqslant n\eta _{-}(\mu ^{2}\mathbf{)}\leqslant \frac{nM}{\sqrt{2}}.
\end{equation*}%
We see that the vertical sizes of the regions $\mathcal{\tilde{A}}_{i}$, $%
i\neq 0$, are bounded by $nM/\sqrt{2}$, which tends to zero for $%
M\rightarrow 0$. This means that all the regions $\mathcal{\tilde{A}}_{i}$, $%
i\neq 0$, squeeze onto the real axis in that limit. Thus, the fakeons with $%
\mathcal{E},M\rightarrow 0$ do not need the domain deformation.

\section{Perturbative unitarity}

\setcounter{equation}{0}\label{uni}

In this section we derive the cutting equations and prove that the fakeon
theories are perturbatively unitary to all orders. We assume that the
Lagrangian is local and Hermitian.

Writing the $S$ matrix as $S=1+iT$, the unitarity relation $SS^{\dagger }=1$%
, which is equivalent to $T-T^{\dag }=iTT^{\dag }$, can be expressed
diagrammatically by means of the so-called cutting equations \cite%
{unitarity,unitaritymio,ACE}, which relate the discontinuities of the
amplitudes to sums of \textquotedblleft cut diagrams\textquotedblright . The
cut diagrams are built with the usual vertices and propagators, plus their
Hermitian conjugates, as well as \textquotedblleft cut
propagators\textquotedblright . The cut propagators play a crucial role,
because they tell us which degrees of freedom are propagated by the theory.
Precisely, they encode the key completeness relation, which allows us to
derive the unitarity equation $SS^{\dagger }=1$ from the cutting equations.
If ghosts are present, the cutting equations are still meaningful, but lead
to a pseudounitarity equation instead of $SS^{\dagger }=1$.

We want to prove that the fakeon models admit a \emph{physical} subspace $V$
of states and unitary cutting equations. This means that, if we project the
initial and final states $|\alpha \rangle ,|\beta \rangle $ onto $V$, only
states $|n\rangle $ belonging to $V$ propagate through the cuts of the
cutting equations. In other words, the completeness relation%
\begin{equation}
\mathds{1}=\sum_{|n\rangle \in V}|n\rangle \langle n|  \label{crel}
\end{equation}%
holds in $V$, so that 
\begin{equation}
|\alpha \rangle ,|\beta \rangle \in V\qquad \Longrightarrow \qquad \langle
\alpha |T|\beta \rangle -\langle \alpha |T^{\dag }|\beta \rangle
=i\sum_{|n\rangle \in V}\langle \alpha |T|n\rangle \langle n|T^{\dag }|\beta
\rangle .  \label{unit}
\end{equation}

Obviously, we cannot demand unitarity for arbitrary complex external
momenta, because the physical momenta are real. Therefore, we derive cutting
equations that hold in a neighborhood $\mathcal{U}_{R}\subset \mathcal{P}$
of the subspace of real momenta and conclude that, thanks to them, the $S$ matrix is
unitary for real (on shell) external momenta. Note that the cutting
equations also hold off shell.

We can assume that the LW scales $M$ are arbitrary and different from zero.
Once perturbative unitarity is proved in that case, it also follows for
evanescent LW scales $\mathcal{E}$, as long as they tend to zero after the
widths $\epsilon$.

The strategy of the proof is as follows. We first derive more general
versions of the cutting equations that hold when the external momenta belong
to the Euclidean region and the widths $\epsilon $ are nonvanishing. Then,
we extend the validity of those equations to $\mathcal{U}_{R}\cap \mathcal{A}%
_{0}$ by analytic continuation and prove that they have the expected,
unitary form in the limit $\epsilon \rightarrow 0$. Third, we
average-continue the generalized cutting equations of $\mathcal{U}_{R}\cap 
\mathcal{A}_{0}$ to $\mathcal{U}_{R}\cap \mathcal{A}_{i}$, $i\neq 0$, at $%
\epsilon \neq 0$. Finally, we show that, in the limit $\epsilon \rightarrow
0 $, the equations have the correct unitary form in every $\mathcal{U}%
_{R}\cap \mathcal{A}_{i}$.

We begin by recalling an important tool that we use in the proof, i.e. the
algebraic cutting equations.

\subsection{Algebraic cutting equations}

\begin{figure}[t]
\begin{center}
\includegraphics[width=14cm]{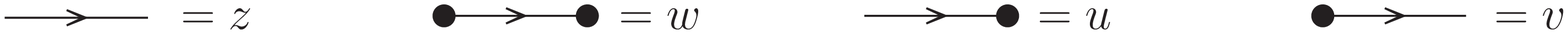}
\end{center}
\caption{ACE propagators}
\label{feybub}
\end{figure}

The algebraic cutting equations \cite{ACE} are particular polynomial
identities associated with Feynman diagrams. Let $\{\sigma _{i}^{+},\tau
_{i}^{+},\sigma _{i}^{-},\tau _{i}^{-}\}$, $i=1,\ldots N$, denote $N$ sets
made of four variables each. An abstract marking, called \textit{polarity}
and specified by the superscripts $+$ or $-$, is assigned to these
variables. We say that $\sigma _{i}^{+}$, $\tau _{i}^{+}$ (resp. $\sigma
_{i}^{-}$, $\tau _{i}^{-}$) are positive (negative) \textit{polar numbers}
and use them to define the \textit{propagators}%
\begin{equation}
z_{i}=\sigma _{i}^{+}+\sigma _{i}^{-},\qquad w_{i}=\tau _{i}^{+}+\tau
_{i}^{-},\qquad u_{i}=\sigma _{i}^{+}+\tau _{i}^{-},\qquad v_{i}=\sigma
_{i}^{-}+\tau _{i}^{+}.  \label{propagators}
\end{equation}

Consider a Feynman diagram $G$ with $I$ internal legs and $V$ vertices. We
may assume that $G$ is connected. Equip the $G$ internal legs with
orientations. We say that a curve is oriented if the orientations of all its
legs are coherent. We say that a loop, i.e. a closed curve, is minimal if it
is not the union of two loops that have a vertex in common.

Assign an independent energy to each internal leg and assume that it flows
according to the leg orientation. Then, impose the energy conservation at
each vertex, with zero energies on the external legs. This leaves $L=I-V+1$
independent energies $e_{1},\ldots e_{L}$. We can arrange the orientations
and the energies so that the flow of each $e_{i}$ defines an oriented
minimal loop and the energy flowing in each internal leg is a linear
combination of $e_{1},\ldots e_{L}$ with coefficients 0 or 1. In this case,
the diagram is said to be oriented.

Build variants $G_{\mathfrak{M}}$ of an oriented diagram $G$ by marking any
number of vertices. We define the \textit{value} $P_{\mathfrak{M}}$ of $G_{%
\mathfrak{M}}$ by means of the following rules. Give value 1 to each
unmarked vertex and value $-1$ to each marked vertex. Assign the propagators
shown in fig. \ref{feybub} to the internal legs of $G_{\mathfrak{M}}$, where
the dots denote the marked vertices. Then, $P_{\mathfrak{M}}$ is the product
of the values associated with the vertices and the propagators.

The algebraic cutting equation associated with $G$ is the polynomial identity%
\begin{equation}
\sum_{\text{markings }\mathfrak{M}}P_{\mathfrak{M}}=\mathcal{P}_{G},
\label{theorem}
\end{equation}%
where $\mathcal{P}_{G}$ is a linear combination of \textit{polarized
monomials}. A polarized monomial is a product of polar numbers, one for each
internal leg, where at least one loop $\gamma $ is polarized. We say that $\gamma$ is polarized if
the polar numbers associated with the legs of $\gamma $ are arranged so
that, moving along $\gamma $, the polarization flips if and only if the leg
orientation flips.

The main virtue of the identity (\ref{theorem}) is that it isolates the
terms (those collected on the right-hand side) that do not contribute to the 
\textit{diagrammatic} cutting equations. Indeed, in typical applications the
polarity of a polar number refers to the position of its poles with respect
to the integration path on the loop energy. A polarized loop is a product of
polar numbers whose poles are all located on the same side. Letting tadpoles
and nontrivial numerators aside, which can be treated with little additional
effort \cite{ACE}, if we apply the residue theorem to perform the integral
on the energy of a polarized loop, the result is zero.

\begin{figure}[t]
\begin{center}
\includegraphics[width=10truecm]{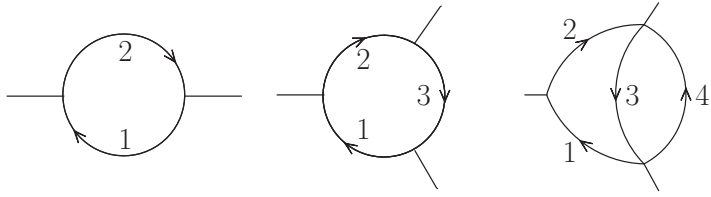}
\end{center}
\caption{Oriented diagrams}
\label{example}
\end{figure}

To give a few examples, consider the diagrams of fig. \ref{example}. The
oriented loops of the third diagram are 123 and 34. Instead, 124 is a
nonoriented loop. Equipped with polar monomials such as $\sigma
_{1}^{+}\sigma _{2}^{+}\tau _{3}^{+}$, $\sigma _{3}^{-}\tau _{4}^{-}$ and $%
\tau _{1}^{+}\sigma _{2}^{+}\sigma _{4}^{-}$, respectively, these loops
become polarized. Examples of polarized monomials for the third diagram are $%
\sigma _{1}^{+}\sigma _{2}^{+}\tau _{3}^{+}\sigma _{4}^{-}$ and $\sigma
_{1}^{+}\sigma _{2}^{+}\sigma _{3}^{-}\tau _{4}^{-}$.

The polynomial identities (\ref{theorem}) associated with the diagrams of
fig. \ref{example} are%
\begin{eqnarray*}
&&z_{1}z_{2}+w_{1}w_{2}-u_{1}v_{2}-v_{1}u_{2}\sim 0, \\
&&z_{1}z_{2}z_{3}-w_{1}w_{2}w_{3}-u_{1}v_{2}z_{3}-z_{1}u_{2}v_{3}-v_{1}z_{2}u_{3}+v_{1}u_{2}w_{3}+w_{1}v_{2}u_{3}+u_{1}w_{2}v_{3}\sim 0,
\\
&&z_{1}z_{2}z_{3}z_{4}-w_{1}w_{2}w_{3}w_{4}-u_{1}v_{2}z_{3}z_{4}-v_{1}z_{2}u_{3}v_{4}-z_{1}u_{2}v_{3}u_{4}
\\
&&\qquad \qquad \qquad \qquad \qquad \qquad \qquad \qquad
+v_{1}u_{2}w_{3}w_{4}+u_{1}w_{2}v_{3}u_{4}+w_{1}v_{2}u_{3}v_{4}\sim 0,
\end{eqnarray*}%
where the polarized monomials on the right-hand sides have been replaced by
zeros, since in the end they do not contribute to the diagrammatic cutting
equations.

The algebraic cutting equations are more general than the usual diagrammatic
cutting equations that are met in physics, in the sense that no particular
assumptions are made about the polar numbers, apart from their polarity
assignments. In the usual applications to quantum field theory, $z_{i}$ are
the ordinary propagators and $w_{i}$ are their complex conjugates. Moreover, 
$u_{i}$ and $v_{i}$ are the cut propagators, i.e. distributions of compact
support, typically theta functions that multiply delta functions. Here it is
not necessarily so. For example, we are free to keep the infinitesimal
widths $\epsilon $ of the Feynman prescription different from zero and
arbitrary. Being able to work at $\epsilon \neq 0$ is crucial to prove the
perturbative unitarity of the fakeon models.

\subsection{Perturbative unitarity of the fakeon models in the Euclidean
region}

In the first step of the proof, we concentrate on the Euclidean region,
which is the region where every linear combination $p=\sum_{i\in I}p_{i}$ of
incoming momenta that appears in formula (\ref{thresh}) satisfies $|\mathrm{%
Re}[p^{0}]|<|\mathbf{p}|$. Clearly, the region is open and nonempty.

\begin{figure}[t]
\begin{center}
\includegraphics[width=8truecm]{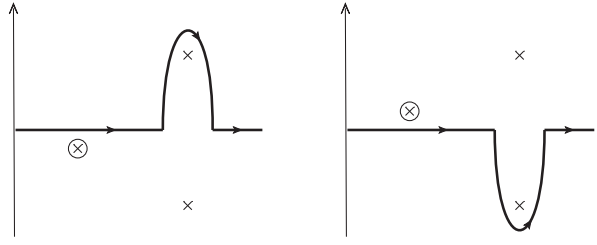}
\end{center}
\caption{Poles of $\protect\sigma ^{+}$ and $\protect\tau ^{-}$}
\label{propw}
\end{figure}

We write the propagator (\ref{propa}) as $\sigma ^{+}+\sigma ^{-}$, where\
the polar numbers $\sigma ^{+}$ and $\sigma ^{-}$ are%
\begin{equation}
\sigma ^{\pm }(p)=\pm \frac{a}{p^{0}\mp \omega _{\epsilon }(\mathbf{p})}+%
\frac{ib^{\pm }}{p^{0}\mp \Omega ^{\pm }(\mathbf{p})}-\frac{ib^{\mp }}{%
p^{0}\mp \Omega ^{\mp }(\mathbf{p})},  \label{spmk}
\end{equation}%
with%
\begin{equation*}
a=\frac{1}{2\omega}\frac{M^{4}}{(m^{2}-\mu ^{2})^{2}+M^{4}}%
,\qquad b^{\pm }=\frac{1}{4\Omega ^{\pm }}\frac{M^{2}}{m^{2}-\mu ^{2}\mp
iM^{2}}.
\end{equation*}%
Observe that $b^{\pm }=(b^{\mp })^{\ast }$, $\Omega ^{\pm }=(\Omega ^{\mp
})^{\ast }$, $\omega=\sqrt{\mathbf{p}^{2}+m^{2}}$ and $a$ is
real. We have replaced $m^{2}-i\epsilon $ with $m^{2}$ in the coefficients $%
a $, $b^{\pm }$, since the limit $\epsilon \rightarrow 0$ is trivial there.
Here and below the complex conjugation denoted with a $\ast $ does not act
on the momenta.

We define $\tau ^{\pm }=-(\sigma ^{\mp })^{\ast }$ and the propagators 
\begin{equation}
z=\sigma ^{+}+\sigma ^{-},\qquad w=\tau ^{+}+\tau ^{-},\qquad u=\sigma
^{+}+\tau ^{-},\qquad v=\sigma ^{-}+\tau ^{+}.  \label{zwuv}
\end{equation}%
Observe that the contributions of the LW poles disappear from the cut
propagators $u$ and $v$, which simplify to%
\begin{equation}
u=\frac{a}{p^{0}-\omega _{\epsilon }(\mathbf{p})}-\frac{a}{p^{0}-\omega
_{\epsilon }^{\ast }(\mathbf{p})},\qquad v=-\frac{a}{p^{0}+\omega _{\epsilon
}(\mathbf{p})}+\frac{a}{p^{0}+\omega _{\epsilon }^{\ast }(\mathbf{p})}.
\label{uv}
\end{equation}%
The limits of these expressions for $\epsilon \rightarrow 0$ are the cut
propagators we expect (apart from an overall factor), i.e.%
\begin{equation}
u\rightarrow -\frac{2i\pi M^{4}}{(m^{2}-\mu ^{2})^{2}+M^{4}}\theta
(p^{0})\delta (p^{2}-m^{2}),\qquad v\rightarrow -\frac{2i\pi M^{4}}{%
(m^{2}-\mu ^{2})^{2}+M^{4}}\theta (-p^{0})\delta (p^{2}-m^{2}).  \label{esf}
\end{equation}%
These results put the physical degrees of freedom on shell and are
independent of the LW poles. It seems that perturbative unitarity may follow
straightforwardly from (\ref{esf}). Unfortunately, this argument is too
naive, for the following reason. 

Recall, from the previous subsection, that the notion of polarity must allow
us to drop the right-hand side of formula (\ref{theorem}). Consider the
positions of the poles of $\sigma ^{\pm }$ and $\tau ^{\pm }$ with respect
to the integration path on $p^{0}$, when the propagators appear in a loop
diagram. We see that the poles of $\sigma ^{+}$ and $\tau ^{+}$ are placed
below the integration path, while those of $\sigma ^{-}$ and $\tau ^{-}$ are
placed above the integration path. Thus, having positive (resp. negative)
polarity means \textquotedblleft having poles placed below (above) the
integration path on the energy\textquotedblright . To take care of this in
the relation $\tau ^{\pm }=-(\sigma ^{\mp })^{\ast }$, we must flip the
integration path accordingly, as shown in fig. \ref{propw}. The left picture
of fig. \ref{propw} shows the $p^{0}$ poles of $\sigma ^{+}$, which are made
of a physical pole and a LW\ pair, while the right picture shows the $p^{0}$
poles of $\tau ^{-}=-(\sigma ^{+})^{\ast }$. If we put the two sets of poles
together, we obtain the cut propagator $u$, which gives fig. \ref%
{selfLWpinch}, where the LW poles must be further displaced till the top
ones, as well as the bottom ones, come to coincide. Clearly, when we do
this, the integration path gets pinched. We call this kind of pinching 
\textit{LW\ selfpinching}, since it does not involve different propagators,
but the poles of the same (cut) propagator. A mirror picture with respect to
the imaginary axis is obtained for $v$. Observe that when $\epsilon $ tends
to zero the standard poles of the cut propagators also pinch the integration
path. We call that pinching \textit{standard} selfpinching. 
\begin{figure}[t]
\begin{center}
\includegraphics[width=5truecm]{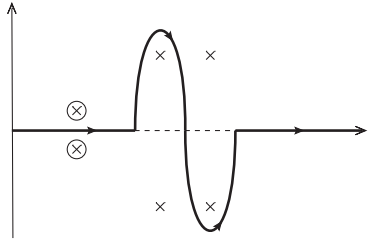}
\end{center}
\caption{Poles of the cut propagator $u$ and LW selfpinching}
\label{selfLWpinch}
\end{figure}

\begin{figure}[t]
\begin{center}
\includegraphics[width=7truecm]{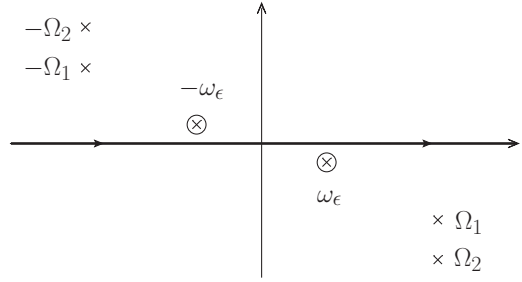}
\end{center}
\caption{Poles of the half propagators (\protect\ref{sigpm})}
\label{propmod}
\end{figure}

To describe the LW selfpinching more clearly, it is convenient to start from
different polar numbers, located in more usual positions, as shown in fig. %
\ref{propmod}. Specifically, we take%
\begin{equation}
\sigma ^{\pm }(p)=\pm \frac{a}{p^{0}\mp \omega _{\epsilon }(\mathbf{p})}+%
\frac{ib^{\pm }}{p^{0}\mp \Omega _{1}(\mathbf{p})}-\frac{ib^{\mp }}{p^{0}\mp
\Omega _{2}(\mathbf{p})},  \label{sigpm}
\end{equation}%
where $\Omega _{1}$, $\Omega _{2}$ have negative imaginary parts, together
with $\tau ^{\pm }=-(\sigma ^{\mp })^{\ast }$. Now the polar number $\sigma
^{+}$ has three poles located in the fourth quadrant, while the polar number 
$\sigma ^{-}$ has three poles located in the second quadrant. For example,
making the $M$ dependence explicit by writing $\Omega ^{\pm }(\mathbf{p},M)=%
\sqrt{\mathbf{p}^{2}+M_{\pm }^{2}}$, we can set $\Omega _{1}(\mathbf{p}%
)=\Omega ^{-}(\mathbf{p},M^{\prime })$ and $\Omega _{2}(\mathbf{p})=\Omega
^{-}(\mathbf{p},M)$ for some real $M^{\prime }\neq M$. For the arguments that
follow, it may also be convenient to pick a different $M^{\prime }$ for
every propagator.

If we keep the definitions (\ref{zwuv}) and take the real axis as the
integration path for the energies, we can derive the algebraic cutting
equations (\ref{theorem}) using the polar numbers (\ref{sigpm}), by applying
the Feynman rules of the previous subsection. When we integrate on the loop
momenta, the right-hand side drops out, which leads to the diagrammatic
cutting equations 
\begin{equation}
G+\bar{G}=-\sum_{\text{proper markings }\mathfrak{M}}G_{\mathfrak{M}},
\label{cuto}
\end{equation}%
where the sum is over the properly marked diagrams $G_{\mathfrak{M}}$, i.e.
the diagrams that contain at least one marked vertex and one unmarked
vertex. The diagram $\bar{G}$ is the one with all marked vertices.

We have taken nonderivative vertices, so far, but the arguments also work
when the vertices are polynomials of the momenta and the free propagators
have nontrivial polynomial numerators. We stress once again that the
equations (\ref{cuto}) that we obtain are more general than the usual
cutting equations, since the widths $\epsilon $ do not need to be small or
tend to zero, but are completely arbitrary.

As long as the polar numbers are (\ref{sigpm}), the Wick rotation is
straightforward. Then, however, the cut propagators do not simplify as in (%
\ref{uv}) and do not reduce to the expected form (\ref{esf}) when $\epsilon
\rightarrow 0$. We must migrate $\Omega _{1}(\mathbf{p})$ to $\Omega ^{+}(%
\mathbf{p},M)$, which is equivalent to complexify $M^{\prime }$ and deform $%
M^{\prime \hspace{0.01in}2}$ continuously into $-M^{2}$. During the
migration, $\Omega _{1}$ crosses the real axis. To keep the algebraic
cutting equations valid, $\Omega _{2}$ cannot cross the integration path,
since the definition of polarity refers to the positions of the poles in $%
p^{0}$ with respect to it. Thus, we have to deform the integration path so
as to avoid the crossing. This operation, applied on $\sigma ^{+}$, leads to
the first picture of fig. \ref{propw}. It leads to the second picture of
fig. \ref{propw} when it is applied to $\tau ^{-}=-(\sigma ^{+})^{\ast }$.
When we apply it to the cut propagators $u$ and $v$, we must take into
account that the LW pair of $\sigma ^{+}$ and the LW pair of $(\sigma
^{+})^{\ast }$ remain on opposite sides of the integration path, which leads
to fig. \ref{selfLWpinch} and its reflection with respect to the imaginary
axis. This is the reason why we cannot drop the LW\ pairs from the
difference $u=\sigma ^{+}-(\sigma ^{+})^{\ast }$ so quickly. First, we have
to make one LW pair cross the integration path. Once it is on the other
side, it does \textquotedblleft annihilate\textquotedblright\ the other
pair. However, the crossing leaves a remnant (the contributions of a pair of
residues), which must be taken into account. To prove perturbative unitarity
we need to show that such a remnant does not contribute to the cutting
equations.

Observe that the crossing only concerns the cut propagators. In uncut
propagators, the migration of the poles $\Omega _{1}(\mathbf{p})$ just
returns the right result, shown in fig. \ref{prop}, and no selfpinching
occurs. For this reason, the left-hand side of the cutting equation (\ref%
{cuto}) goes directly to its correct, final form. Only the right-hand side
needs a detailed analysis.

Consider a properly marked diagram $G_{\mathfrak{M}}$. Assume that the cut
propagators are $n+1$ and depend on $n$ loop momenta (the most general case
being a straightforward generalization of this one). Each cut propagator $%
u=\sigma ^{+}+\tau ^{-}$ and $v=\sigma ^{-}+\tau ^{+}$ receives
contributions from LW\ selfpinchings and standard selfpinchings. We
decompose $G_{\mathfrak{M}}$ as a sum of terms where each cut propagator
involves either of the two. We analyze such terms one by one, starting from
the terms that involve only LW selfpinchings.

Integrate on the $n$ loop energies $k_{i}^{0}$ by means of the residue
theorem and take $M^{\prime \hspace{0.01in}2}\rightarrow -M^{2}$ in $n$ cut
propagators. This operations give $n$ conditions of the form $k_{i}^{0}=%
\tilde{\omega}_{i}(\mathbf{k}_{i})$, which eliminate the loop energies $%
k_{i}^{0}$. At this point, the contribution of the LW selfpinching due the
last cut propagator has the form 
\begin{equation}
\frac{1}{D_{\text{pinch}}^{+}}-\frac{1}{D_{\text{pinch}}^{-}},  \label{diffa}
\end{equation}%
where $D_{\text{pinch}}^{\pm }$ are deformed versions of the denominators $%
D_{\text{pinch}}$ of equation (\ref{pinchgen}). The deformations depend on $%
M^{\prime }$ and are such that $D_{\text{pinch}}^{\pm }\rightarrow D_{\text{%
pinch}}$ when $M^{\prime \hspace{0.01in}2}\rightarrow -M^{2}$. Moreover,
they make $D_{\text{pinch}}^{\pm }$ vanish on opposite sides of the
integration path.

After the integrations on $k_{i}^{0}$, the integration path has actually
disappeared, so formula (\ref{diffa}) can be read as it stands. When we
finalize the migration of $\Omega ^{-}(\mathbf{p},M^{\prime })$ into $\Omega
^{+}(\mathbf{p},M)$ by taking the limit $M^{\prime \hspace{0.01in}%
2}\rightarrow -M^{2}$, the difference (\ref{diffa}) gives zero, because we
are working in the Euclidean region, where the loop space momenta are integrated on their natural real domains and the condition $D_{\text{pinch}}=0$
has no solutions. We recall that, indeed, $D_{\text{pinch}}=0$ is the
condition for having a LW pinching, which defines the regions $\mathcal{%
\tilde{A}}_{i}$, $i\neq 0$.

Now, consider the terms where only standard selfpinchings occur. Those are
the expected terms, the only ones that should survive at the very end.
Indeed, the differences (\ref{diffa})\ give (\ref{uv}) in this case.

Finally, consider the mixed selfpinching, i.e. the terms where the
contributions of some cut propagators come from LW selfpinchings and those
of other cut propagators come from standard selfpinchings. Recall that the
LW selfpinching occurs when we complete the migration of $\Omega ^{-}(%
\mathbf{p},M^{\prime })$ into $\Omega ^{+}(\mathbf{p},M)$ by taking $%
M^{\prime \hspace{0.01in}2}\rightarrow -M^{2}$. Instead, the standard
selfpinching occurs when we take $\epsilon \rightarrow 0$. If we are willing
to let the widths $\epsilon $ disappear at the end, the argument used for
the terms with only LW selfpinchings can be applied with straightforward
modifications and leads to the conclusion that the contributions of the
mixed selfpinchings vanish in the limit $\epsilon \rightarrow 0$. For
various arguments that follow, however, it is necessary to keep $\epsilon
\neq 0$. There, we have generalized cutting equations that contain extra
contributions, which must be taken into account for the extension of the
proof beyond the Euclidean region. For example, consider the case where the
contributions of the first $n$ cut propagators come from LW selfpinchings
and those of the last cut propagator come from a standard selfpinching with
mass $m$. We integrate on the $n$ loop energies as before and complete the
migrations $M^{\prime \hspace{0.01in}2}\rightarrow -M^{2}$. At the last
step, we obtain an integrand proportional to an expression of the form (\ref%
{diffa}), where the denominators $D_{\text{pinch}}^{\pm }$ are equal to (\ref%
{deno}) with $r=1$ and imaginary parts $\mp i\epsilon $ attached to the
squared mass $m^{2}$. Clearly, (\ref{diffa}) does not vanish in this case
until we take $\epsilon \rightarrow 0$.

Summarizing, the expected, unitary cutting equations hold in the Euclidean
region for $\epsilon \rightarrow 0$. The cut propagators can be effectively
replaced by (\ref{esf}) in that limit and the LW degrees of freedom do not
propagate through the cuts. Moreover, generalized cutting equations hold at $%
\epsilon \neq 0$.

\subsection{Perturbative unitarity in the other regions}

The next step is to extend the validity of the generalized cutting equations
by analytic continuation from the Euclidean region to the intersection $%
\mathcal{U}_{R}\cap \mathcal{A}_{0}$. Then we have to reach the other
regions $\mathcal{U}_{R}\cap \mathcal{A}_{i}$, $i\neq 0$, by means of the average
continuation. In both cases, we must prove that the generalized cutting
equations reduce to the expected, unitary cutting equations in the limit $%
\epsilon \rightarrow 0$. We assume that the masses are arranged so that the
LW thresholds are all distinct.

We have seen that the generalized cutting equations in the Euclidean region
have corrections $\mathcal{C}(p,\epsilon )$ for $\epsilon \neq 0$, due the
mixed selfpinchings, where $p$ are the incoming external momenta. 
The reason why they vanish for $\epsilon \rightarrow 0$ is that $D_{\text{%
pinch}}$ never vanishes in the Euclidean region.

The first extension away from the Euclidean region is straightforward. At $%
\epsilon \neq 0$ the standard branch points are displaced from the real
axis. Moreoveor, we know that we can deform the integration domain on the
loop space momenta so as to avoid the LW pinchings everywhere in $\mathcal{A}%
_{0}$. Once we do that, we can analytically continue the generalized cutting
equation (\ref{cuto}) from the Euclidean region to $\mathcal{U}_{R}\cap 
\mathcal{A}_{0}$ by keeping $\epsilon \neq 0$ and moving along the real
axis. Then, the corrections $\mathcal{C}(p,\epsilon )$ still vanish when we
take the limit $\epsilon \rightarrow 0$, because $D_{\text{pinch}}$ never
vanishes.

When we attempt to analytically continue the cutting equation (\ref{cuto})
above an LW threshold $P$, we find that it cannot be done in a unique way.
Averaging the two independent ways of doing it, we can prove perturbative
unitarity in the regions $\mathcal{U}_{R}\cap \mathcal{A}_{i}$, $i\neq 0$.

Specifically, we make the two domain deformations $\mathcal{D}_{\mathbf{k},%
\mathbf{q}}\rightarrow \mathcal{D}_{\mathbf{k},\mathbf{q}}^{+\text{def}}$
and $\mathcal{D}_{\mathbf{k},\mathbf{q}}\rightarrow \mathcal{D}_{\mathbf{k},%
\mathbf{q}}^{-\text{def}}$ explained at the end of section \ref{avediffe}.
Applying the deformations on the entire cutting equation (\ref{cuto}), we
obtain two deformed versions of it.

In the case of the deformation $\mathcal{D}_{\mathbf{k},\mathbf{q}%
}\rightarrow \mathcal{D}_{\mathbf{k},\mathbf{q}}^{+\text{def}}$, we denote
the deformed versions of the diagrams $G$, $\bar{G}$ and $G_{\mathfrak{M}}$
by $\mathcal{J}_{+}$, $\mathcal{\bar{J}}_{+}$ and $\mathcal{J}_{+\mathfrak{M}%
}$, respectively. In the case of the deformation $\mathcal{D}_{\mathbf{k},%
\mathbf{q}}\rightarrow \mathcal{D}_{\mathbf{k},\mathbf{q}}^{-\text{def}}$,
we denote them by $\mathcal{J}_{-}$, $\mathcal{\bar{J}}_{-}$ and $\mathcal{J}%
_{-\mathfrak{M}}$. In each case, we obtain an integral representation of the
cutting equation (\ref{cuto}) in some interval $\mathcal{I}$ of the real
axis above $P$ and we can reach $\mathcal{I}$ by analytic continuation from
the Euclidean region without encountering LW\ pinchings. Since $D_{\text{%
pinch}}$ never vanishes in $\mathcal{I}$, the corrections $\mathcal{C}%
(p,\epsilon )$ still vanish for $\epsilon \rightarrow 0$.

Note that the left-hand sides $\mathcal{J}_{\pm }+\mathcal{\bar{J}}_{\pm }$
of the deformed cutting equations are no longer real, because the integral
representations of $\mathcal{J}_{\pm }$ and $\mathcal{\bar{J}}_{\pm }$ have
the same (complex)\ deformed domains $\mathcal{D}_{\mathbf{k},\mathbf{q}%
}^{\pm \text{def}}$. By construction we have $\mathcal{\bar{J}}_{\pm }=(%
\mathcal{J}_{\mp })^{\ast }$.

When we average the two deformed cutting equations, we obtain the cutting
equation that holds above the LW threshold. The average of the left-hand
sides gives 
\begin{equation*}
\frac{1}{2}(\mathcal{J}_{+}+\mathcal{\bar{J}}_{+})+\frac{1}{2}(\mathcal{J}%
_{-}+\mathcal{\bar{J}}_{-})=\frac{1}{2}(\mathcal{J}_{+}+\mathcal{J}_{-})+%
\frac{1}{2}(\mathcal{\bar{J}}_{+}+\mathcal{\bar{J}}_{-}),
\end{equation*}%
where $(\mathcal{J}_{+}+\mathcal{J}_{-})/2$ is the average continuation of $%
G $ and $(\mathcal{\bar{J}}_{+}+\mathcal{\bar{J}}_{-})/2$ is the average
continuation of $\bar{G}$. The average of the right-hand sides has the
expected form for $\epsilon \rightarrow 0$, since the contributions $%
\mathcal{C}(p,\epsilon )$ drop out in that limit.

The conclusion holds in the neighborhood of every $\mathcal{I\subset U}%
_{R}\cap \mathcal{A}_{i}$, so it also holds in the whole $\mathcal{U}%
_{R}\cap \mathcal{A}_{i}$. Applying this procedure to each LW\ threshold at
a time, we reach every $\mathcal{U}_{R}\cap \mathcal{A}_{i}$, $i\neq 0$.
When anomalous thresholds are met, there are multiple ways to circumvent
them, which correspond to multiple options for the deformations, as
described at the end of section \ref{thedomdef}. Each option can be used to
average-continue the cutting equations as described above. The corrections $%
\mathcal{C}(p,\epsilon )$ vanish for $\epsilon \rightarrow 0$ in every case.

In the end, the cutting equations have the expected unitary form in all the
regions $\mathcal{A}_{i}$ for $\epsilon \rightarrow 0$. This concludes the
proof that the fakeon models are perturbatively unitary to all orders. Note
that it would be much more difficult to make the extension to $\mathcal{A}%
_{i}$, $i\neq 0$, using the nonanalytic Wick rotation. This shows once more
the power of the average continuation, a very simple operation that allows
us to make a number of manipulations that otherwise would be very cumbersome.

\subsection{Remarks}

Before concluding this section, we comment on the resummation of the
perturbative series and its effects on the unitarity equation $SS^{\dagger
}=1$. We recall that the LW\ poles of the free propagators (\ref{propa}) are
located symmetrically with respect to the real axis. This is important for
the proof of perturbative unitarity, because the contributions of complex
conjugate LW poles compensate each other. However, the exact two-point
functions may lose the symmetry just mentioned, because the resummations may
give widths to the standard poles and the LW poles, and change their masses.
This is no source of concern, because that symmetry, which is helpful to see
unitarity at the perturbative level, plays no role after the resummations.

Once we have derived the diagrammatic cutting equations (\ref{cuto}) and
projected the external states onto $V$, we have the completeness relation (%
\ref{crel}) and the unitarity equations (\ref{unit}). At a first stage, let
us ignore the resummations that affect the standard poles and concentrate on
those that affect the LW poles. Then the states of $V$ stay the same and the
unitarity equations (\ref{unit}) remain valid. These types of resummations
just act internally to the correlation functions associated with $\langle
\alpha |T|\beta \rangle $, $\langle \alpha |T^{\dag }|\beta \rangle $, $%
\langle \alpha |T|n\rangle $ and $\langle n|T^{\dag }|\beta \rangle $. At a
second stage, we perform the resummations that affect the standard poles.
Some physical particles may acquire widths and decay, and so disappear from
the physical spectrum at very large distances. Since they still propagate
through the cuts of the cutting equations, the $S$ matrix is no longer
unitary in a strict sense, although it remains perturbatively unitary.

In other words, when we resum the perturbative expansion, the LW sector does
not affect unitarity. Yet, some physical poles may get nonvanishing widths,
pretty much like the muon in the standard model. In this respect, the fakeon
models behave as an ordinary model.

If the Lagrangian is Hermitean, the results of the next section ensure that
its renormalization is also Hermitean, so the denominators of the
renormalized propagators obtained by including the counterterms still have
the structure displayed in formula (\ref{propa}), with pairs of complex
conjugate poles, besides the physical poles.

\section{Renormalizability}

\setcounter{equation}{0}\label{renormalization}

Commonly, higher-derivative theories are thought to have an enhanced power
counting, because the propagators fall off more rapidly at high energies.
However, the usual rules of power counting just work in Euclidean space,
while in Minkowski spacetime it is much more difficult to have control on
the ultraviolet behaviors of the Feynman diagrams. Everything is fine if the
Minkowski formulation of the theory is analytically equivalent to the Wick
rotated Euclidean one, which happens for example when the free propagators
just have poles on the real axis. A fakeon model does not have this
property, to the extent that the Minkowski version is plagued by nonlocal,
non-Hermitian counterterms \cite{ugo}. At the same time, we know that the
Wick rotation of the Euclidean version of a fakeon model is not analytic
everywhere, so we have reasons to worry that the nice renormalizability
properties of the Euclidean version may not be fully inherited by the
nonanalytically Wick rotated theory.

In this section we overcome these worries, by proving that the
renormalization of a fakeon\ model is still local and actually coincides
with the one of its Euclidean version. We give two arguments, the first one
based on the average continuation and the second one based on the
nonanalytic Wick rotation.

The first argument is straightforward. Once we have subtracted the
divergences of the Euclidean theory, the amplitudes are convergent in the
Euclidean region. We know that we can unambiguously reach every other region from there.
The analytic continuation of a convergent function is obviously convergent.
The same holds for the average continuation, which is made of two analytic
continuations. This implies that the amplitudes are fully convergent in
every analytic region $\mathcal{A}_{i}$.

The second argument requires a bit more work. The rules of power counting of
the Euclidean theory trivially extend from the Euclidean region to the main
region $\mathcal{A}_{0}$, since the Wick rotation is analytic there. So, we
just need to concentrate on the other regions $\mathcal{A}_{i}$, $i\neq 0$.
Let us start from the regions $\mathcal{\tilde{A}}_{i}$, $i\neq 0$, which
are defined as the solutions of the conditions $D_{\text{pinch}}=0$ with
real loop space momenta, $D_{\text{pinch}}$ being given by (\ref{deno}). As
we know, the relative sign in front of the frequencies of (\ref{deno}) is
necessarily positive, otherwise no pinching occurs. Assume that the external
momenta $p$ belong to a compact connected open subset $\mathcal{S}%
_{p}\subset \mathcal{P}$ that contains an open subset of the Euclidean
region. Formula (\ref{deno}) makes it clear that the condition $D_{\text{%
pinch}}=0$ cannot be satisfied in $\mathcal{S}_{p}$ for arbitrarily large $|%
\mathbf{k}_{i}|$ and $|\mathbf{q}_{j}|$. Thus, the solution identifies a
compact subset $\mathcal{C}_{\mathbf{k},\mathbf{q}}$ of the domain $\mathcal{%
D}_{\mathbf{k},\mathbf{q}}$ of the loop space momenta.

Recall that the loop energies $k_{i}^{0}$ are gone after applying the
residue theorem. Split the integral on $\mathcal{D}_{\mathbf{k},\mathbf{q}}$
as the sum of the integral on a compact subset $\mathcal{C}_{\mathbf{k},%
\mathbf{q}}^{\prime }\supset \mathcal{C}_{\mathbf{k},\mathbf{q}}$ plus the
integral on $\mathcal{D}_{\mathbf{k},\mathbf{q}}\backslash \mathcal{C}_{%
\mathbf{k},\mathbf{q}}^{\prime }$. Clearly, the integral on $\mathcal{C}_{%
\mathbf{k},\mathbf{q}}^{\prime }$ is not interested by ultraviolet
divergences. On the other hand, the integral on $\mathcal{D}_{\mathbf{k},%
\mathbf{q}}\backslash \mathcal{C}_{\mathbf{k},\mathbf{q}}^{\prime }$ may be
ultraviolet divergent, but it is not interested by the LW pinching. This
means that it admits an analytic Wick rotation, which makes its ultraviolet
divergences equal to those of its Euclidean version. Observe that the
Euclidean loop integral is reachable analytically while remaining inside $%
\mathcal{S}_{p}$, since $\mathcal{S}_{p}$ is chosen to contain an open
subset of the Euclidean region. Thus, once the (Euclidean) divergences and
subdivergences are subtracted, the loop integral is convergent in $\mathcal{S%
}_{p}$. Since $\mathcal{S}_{p}$ is arbitrary, the subtracted integral is
convergent everywhere in $\mathcal{P}$.

So far, the integration domain $\mathcal{D}_{\mathbf{k},\mathbf{q}}$ is
still undeformed, because we have been working in the regions $\mathcal{%
\tilde{A}}_{i}$. Now we have to perform the domain deformation to go from
the regions $\mathcal{\tilde{A}}_{i}$ to the regions $\mathcal{A}_{i}$. We
can make it so that the deformed $\mathcal{C}_{\mathbf{k},\mathbf{q}}$
remains always compact. Applying the argument above to every deformed $%
\mathcal{D}_{\mathbf{k},\mathbf{q}}$, we see that the final result is
convergent in every region $\mathcal{A}_{i}$.

We conclude that the nonanalyticity of the Wick rotation does not conflict
with the renormalization of the fakeon models, which coincides with the
renormalization of their Euclidean versions. In particular, the locality of
counterterms and the usual rules of power counting hold. This proves that
the fakeon models that are renormalizable do reconcile unitarity and
renormalizability.

\section{Conclusions}

\setcounter{equation}{0}\label{conclusions}

In this paper we have studied the fakeon models, which contain ordinary
physical particles and fakeons, i.e. fake degrees of freedom. An important
subclass are the Lee-Wick models, which have higher derivatives. Fakeons can
also be introduced without higher derivatives, by means of a suitable
quantization prescription.

Formulating the models by nonanalytically Wick rotating their Euclidean
versions, we have shown that they are consistent to all orders. In
particular, we have studied the LW pinching and the domain deformation in
arbitrary diagrams.

The $S$ matrix of the fakeon models is regionwise analytic. Different
analytic regions $\mathcal{A}_{i}$ are related by the average continuation,
a powerful operation that allows us to simplify numerous derivations. The
average continuations of various functions that are frequently met in four,
three and two dimensions have been computed and compared numerically to the
results of the nonanalytic Wick rotation, confirming that the two operations
give the same result.

We have proved that the fakeon models are perturbatively unitary to all
orders. The strategy of the proof was to first use the algebraic cutting
equations to derive generalized versions of the diagrammatic cutting
equations that hold in the Euclidean region at $\epsilon \neq 0$. Then we
have shown that the equations can be analytically continued to the main
analytic region $\mathcal{A}_{0}$ and average-continued to the other
analytic regions $\mathcal{A}_{i}$, $i\neq 0$. Finally, we have proved that
they reduce to the expected, unitary cutting equations when the widths $%
\epsilon $ tend to zero.

Another good property of the fakeon models is that they have the same
renormalization as their Euclidean versions have. This makes them viable
candidates to explain quantum gravity. We recall that while the LW models of
quantum gravity \cite{LWgrav2,LWgravmio} are superrenormalizable, the fakeon
models of quantum gravity can be strictly renormalizable \cite{LWgravmio}.
At present, the best candidate to explain quantum gravity is a fakeon theory
in four dimensions whose Lagrangian density contains the Hilbert-Einstein
term $R$, the cosmological term and the terms $R_{\mu \nu }R^{\mu \nu }$, $%
R^{2}$ \cite{LWgravmio}. It is the unique model whose gauge coupling is
dimensionless. It has all the features we expect apart from one: a
nonvanishing cosmological constant, which may predict a small unitarity
anomaly in the universe. The classical action of this theory coincides with the one
considered in refs. \cite{stelle} and more recently refs. \cite{agravity},
but its quantization and physical predictions are completely
different, because the would-be ghosts have been replaced by the fakeons.
Strictly unitary superrenormalizable models can also be built \cite%
{LWgravmio}, but their features makes them less realistic. In the end, the
fakeon models have all the features that we require to include them into
the set of the physically acceptable theories.

\vskip 12truept \noindent {\large \textbf{Acknowledgments}}

\vskip 2truept

We are grateful to U. Aglietti, L. Bracci and M. Piva for helpful
discussions.


\begin{thebibliography}{99}
\bibitem{leewick} T.D. Lee and G.C. Wick, Negative metric and the unitarity
of the S-matrix, Nucl. Phys. B 9 (1969) 209.

\bibitem{LWqed} T.D. Lee and G.C. Wick, Finite theory of quantum
electrodynamics, Phys. Rev. D 2 (1970) 1033.

\bibitem{CLOP} R.E. Cutkosky, P.V. Landshoff, D.I. Olive, and J.C.
Polkinghorne, A non-analytic S-matrix, Nucl. Phys. B 12 (1969) 281.

\bibitem{lee} T.D. Lee, A relativistic complex pole model with indefinite
metric, in \textit{Quanta: Essays in Theoretical Physics Dedicated to Gregor
Wentzel} (Chicago University Press, Chicago, 1970), p. 260.

\bibitem{nakanishi} N. Nakanishi, Lorentz noninvariance of the complex-ghost
relativistic field theory, Phys. Rev. D 3, 811 (1971).

\bibitem{grinstein} B. Grinstein, D. O'Connell and M.B. Wise, Causality as
an emergent macroscopic phenomenon: The Lee-Wick O(N) model, Phys. Rev. D 79
(2009) 105019 and \href{http://arxiv.org/abs/0805.2156}{arXiv:0805.2156}
[hep-th].

\bibitem{LWformulation} D. Anselmi and M. Piva, A new formulation of
Lee-Wick quantum field theory, J. High Energy Phys. 06 (2017) 066, \href{http://renormalization.com/17a1/}%
{17A1 Renormalization.com} and \href{http://arxiv.org/abs/1703.04584}{%
arXiv:1703.04584} [hep-th].

\bibitem{LWunitarity} D. Anselmi and M. Piva, Perturbative unitarity of
Lee-Wick models, Phys. Rev. D 96 (2017) 045009, \href{http://renormalization.com/17a2/}%
{17A2 Renormalization.com} and \href{http://arxiv.org/abs/1703.05563}{%
arXiv:1703.05563} [hep-th].

\bibitem{unitarity} R.E. Cutkosky, Singularities and discontinuities of
Feynman amplitudes, J. Math. Phys. (NY) 1 (1960) 429;

M. Veltman, Unitarity and causality in a renormalizable field theory with
unstable particles, Physica 29 (1963) 186.

\bibitem{unitaritymio} D. Anselmi, Aspects of perturbative unitarity, Phys.
Rev. D 94 (2016) 025028, \href{http://renormalization.com/16a1/}{16A1
Renormalization.com} and \href{http://arxiv.org/abs/1606.06348}{%
arXiv:1606.06348} [hep-th].

\bibitem{ACE} D. Anselmi, Algebraic cutting equations, \href{http://renormalization.com/16a3/}%
{16A3 Renormalization.com} and \href{http://arxiv.org/abs/1612.07148}{%
arXiv:1612.07148} [hep-th].

\bibitem{thooft} G. 't Hooft, Renormalization of massless Yang-Mills fields,
Nucl. Phys. B 33 (1971) 173;

G. 't Hooft, Renormalizable Lagrangians for massive Yang-Mills fields, Nucl.
Phys. B 35 (1971) 167.

\bibitem{stelle} K.S. Stelle, Renormalization of higher derivative quantum
gravity, Phys. Rev. D 16 (1977) 953;

J. Julve, M. Tonin, \textquotedblleft Quantum gravity with higher derivative
terms,\textquotedblright\ Nuovo Cim. B 46 (1978) 137;

E.S. Fradkin, A.A. Tseytlin, \textquotedblleft Renormalizable asymptotically
free quantum theory of gravity,\textquotedblright\ Nucl. Phys. B 201 (1982)
469;

I. G. Avramidi and A. O. Barvinsky, \textquotedblleft Asymptotic freedom in
higher derivative quantum gravity,\textquotedblright\ Phys. Lett. B 159
(1985) 269.

\bibitem{LWgravmio} D. Anselmi, On the quantum field theory of the
gravitational interactions, J. High Energy Phys. 06 (2017) 086, \href{http://renormalization.com/17a3/}%
{17A3 Renormalization.com} and \href{http://arxiv.org/abs/1704.07728}{%
arXiv:1704.07728} [hep-th].

\bibitem{LWstandardM} B. Grinstein, D. O'Connell, and M.B. Wise, The
Lee-Wick standard model, Phys. Rev. D77 (2008) 025012 and \href{http://arxiv.org/abs/0704.1845}%
{arXiv:0704.1845} [hep-ph];

C.D. Carone and R.F. Lebed, Minimal Lee-Wick extension of the standard
model, Phys. Lett. B668 (2008) 221 and \href{http://arxiv.org/abs/0806.4555}{%
arXiv:0806.4555} [hep-ph];

J.R. Espinosa and B. Grinstein, Ultraviolet properties of the Higgs sector
in the Lee-Wick standard model, Phys. Rev. D83 (2011) 075019 and \href{http://arxiv.org/abs/1101.5538}%
{arXiv:1101.5538} [hep-ph];

C.D. Carone and R.F. Lebed, A higher-derivative Lee-Wick standard model,
JHEP 0901 (2009) 043 and \href{http://arxiv.org/abs/0811.4150}{%
arXiv:0811.4150} [hep-ph].

\bibitem{LWunification} B. Grinstein and D. O'Connell, One-Loop
Renormalization of Lee-Wick Gauge Theory, Phys. Rev. D78 (2008) 105005 and 
\href{http://arxiv.org/abs/0801.4034}{arXiv:0801.4034} [hep-ph];

C. D. Carone, Higher-derivative Lee-Wick unification, Phys. Lett. B677
(2009) 306,

and \href{http://arxiv.org/abs/0904.2359}{arXiv:0904.2359} [hep-ph].

\bibitem{LWgrav} E. Tomboulis, 1/N expansion and renormalization in quantum
gravity, Phys. Lett. B 70 (1977) 361;

E. Tomboulis, Renormalizability and asymptotic freedom in quantum gravity,
Phys. Lett. B 97 (1980) 77.

\bibitem{LWgrav2} 

Shapiro and L. Modesto, Superrenormalizable quantum gravity with complex
ghosts, Phys. Lett. B755 (2016) 279-284 and \href{http://arxiv.org/abs/1512.07600}%
{arXiv:1512.07600} [hep-th];

L. Modesto, Super-renormalizable or finite Lee--Wick quantum gravity, Nucl.
Phys. B909 (2016) 584 and \href{http://arxiv.org/abs/1602.02421}{%
arXiv:1602.02421} [hep-th].

\bibitem{ugo} U.G. Aglietti and D. Anselmi, Inconsistency of Minkowski
higher-derivative theories, Eur. Phys. J. C 77 (2017) 84, \href{http://renormalization.com/16a2/}%
{16A2 Renormalization.com} and \href{http://arxiv.org/abs/1612.06510}{%
arXiv:1612.06510} [hep-th].

\bibitem{infred} F. Bloch and A. Nordsieck, Note on the radiation field of
the electron, Phys. Rev. 52 (1937) 54;

T. Kinoshita, Mass singularities of Feynman amplitudes, J. Math. Phys. 3
(1962) 650;

T. D. Lee and M. Nauenberg, Degenerate systems and mass singularities, Phys.
Rev. 133 (1964) B1549;

S. Weinberg, Infrared photons and gravitons, Phys. Rev. 140 (1965) B516.

\bibitem{anomalousthresh} For details, see R.J. Eden, P.V. Landshoff, D.I.
Olive and J.C. Polkinghorne, \textit{The Analytic S-Matrix}, Cambridge
University Press, Cambridge, UK, 1966.

\bibitem{agravity} A. Salvio and A. Strumia, Agravity up to infinite energy, Eur. Phys. C 78 (2018) 124 and
\href{http://arxiv.org/abs/1705.03896}{arXiv:1705.03896} [hep-th];

A. Salvio and A. Strumia, Agravity, J. High Energ. Phys. 06 (2014) 80 and 
\href{http://arxiv.org/abs/1403.4226}{arXiv:1403.4226} [hep-ph].
\end{thebibliography}
\end{document}